\documentclass[11pt]{article}
\usepackage[marginratio=1:1,height=600pt,width=460pt,tmargin=100pt]{geometry}

\usepackage{authblk}

\usepackage{hyperref}
\usepackage{url}

\usepackage[utf8]{inputenc} %
\usepackage[T1]{fontenc}    %
\usepackage{hyperref}       %
\usepackage{url}            %
\usepackage{booktabs}       %
\usepackage{amsfonts}       %
\usepackage{nicefrac}       %
\usepackage{microtype}      %
\usepackage{xcolor}         %

\usepackage{url}
\usepackage{multicol}
\usepackage{microtype}
\usepackage{graphicx}
\usepackage{tablefootnote}
\usepackage{array}
\usepackage{booktabs} 
\usepackage{cprotect}
\usepackage{wrapfig}
\usepackage{fancyvrb}
\usepackage{soul}
\usepackage{indentfirst}
\usepackage{multirow}
\usepackage{enumitem}
\usepackage{mathtools}
\usepackage{amsmath, amsthm, amssymb}
\usepackage{mathtools}
\usepackage{times}
\usepackage{diagbox}
\usepackage{xcolor} 
\usepackage{color}
\usepackage{mwe}
\usepackage{algorithmic}
\usepackage{algorithm}
\usepackage{bbm}
\usepackage{natbib}
\usepackage{forloop}
\usepackage{url}
\usepackage{multicol}
\usepackage{microtype}
\usepackage{amsthm}
\usepackage{array}
\usepackage{booktabs} %
\usepackage{makecell}
\usepackage{indentfirst}
\usepackage{amsfonts}
\usepackage{physics}
\usepackage{mathdots}
\usepackage{yhmath}
\usepackage{cancel}
\usepackage{hyperref}
\usepackage{tabularx}
\usepackage{extarrows}
\usepackage{lastpage}
\usepackage{bm}

\usepackage{tabularx}
\usepackage{subfigure}

\theoremstyle{plain}
\newtheorem{thm}{Theorem}[section]

\newtheorem{lem}[thm]{Lemma}
\newtheorem{cor}[thm]{Corollary}

\theoremstyle{definition}
\newtheorem{defn}[thm]{Definition}
\newtheorem{assumption}[thm]{Assumption}
\theoremstyle{remark}
\newtheorem{remark}[thm]{Remark}

\renewcommand{\abs}[1]{\left\vert#1\right\vert}

\newcommand{\Real}{\mathbb R}

\def\argmin{\mathop{\rm argmin}}

\def\E{\mathbb{E}}
\def\tr{\mbox{tr}}

\newcommand{\Dc}{\mathcal{D}}

\newcommand{\Fc}{\mathcal{F}}

\newcommand{\Ic}{\mathcal{I}}

\newcommand{\Kc}{\mathcal{K}}

\newcommand{\arlcell}[2]{\shortstack{\rule{0pt}{2.7ex}#1\\{\small (#2)}}}

\def\1v{\mathbf 1}
\def\0v{\mathbf 0}

\def\1{\mathbbm{1}}
\def\P{\mathbb{P}}

\begin{document}

\title{Multi-rank Subspace Change-point Detection with Application in 
Monitoring Robotic Swarms}
\date{}

\author[1]{Jonghyeok Lee}
\author[1]{Yao Xie}
\author[2]{Youngser Park}
\author[4]{Jason Hindes}
\author[4]{Ira B. Schwartz}
\author[2,3]{Carey Priebe}
\affil[1]{{\small H. Milton Stewart School of Industrial and Systems Engineering, Georgia Institute of Technology, Atlanta, GA 30332}}
\affil[2]{{\small Center for Imaging Science, Johns Hopkins University, Baltimore, MD 21218}}
\affil[3]{{\small Department of Applied Mathematics and Statistics, Johns Hopkins University, Baltimore, MD 21218}}
\affil[4]{{\small U.S. Naval Research Laboratory, Code 6792, Washington, DC 20375}}

\maketitle

\begin{abstract}
We study real-time detection of low-rank changes in the covariance structure of high-dimensional streaming data, motivated by robotic swarm monitoring. Building on the spiked covariance model, we propose the Multi-rank Subspace-CUSUM (MRS-C) procedure, which extends classical CUSUM by tracking projection energy onto an estimated signal subspace. We analyze the immediate-change expected detection delay (EDD), deriving closed-form choices of the window size and drift parameter that minimize the leading-order asymptotic EDD approximation. We further establish an oracle-relative asymptotic efficiency result, with an explicit efficiency constant that depends on heterogeneity in spike strengths. When the signal rank is unknown, we propose a practical parallel procedure. Simulations and robotic swarm-behavior data illustrate robustness and effectiveness.
\end{abstract}

\section{INTRODUCTION}

Detecting abrupt changes in high-dimensional streaming data is a fundamental problem in many modern applications, from signal processing to finance and quality control. In many applications, an important class of changes involves the covariance structure of multivariate observations. Such structural changes often reflect the emergence or switching of correlated patterns among multiple sources. For example, the onset of a critical event in a sensor array may induce a common low-dimensional signal across sensors, shifting the covariance matrix from a baseline structure of nearly identity to one with a few dominant eigenvalues. Likewise, in swarm behavior monitoring, the collective motion of agents can lie in a low-dimensional subspace, so a sudden change in behavior manifests as a low-rank change in the observation covariance. Detecting these covariance shifts rapidly and reliably is crucial for a timely response in such systems.

The \emph{spiked covariance} model \citep{johnstone2001distribution} provides a tractable framework for capturing low-dimensional changes in covariance. It assumes that most of the variability is concentrated in a small number of principal components, making it particularly well suited to characterizing emerging signal subspaces. In this model, the covariance matrix consists of an identity baseline plus a low-rank component capturing the dominant subspace. Prior work has developed sequential detectors for rank-1 spike changes, exploiting either the largest sample eigenvalue \citep{xie2020sequential} or an estimated leading eigenvector to construct CUSUM-type procedures \citep{xie2018first}. However, real-world changes often involve multiple concurrent spikes of moderate strength, motivating a generalization to the multi-rank case. To address this, we extend the change-point detection framework to accommodate a general spiked-covariance setting with an unknown multidimensional signal subspace.

The contributions are summarized as follows:

\begin{itemize}
\item We propose the \emph{multi-rank Subspace-CUSUM} algorithm, which detects covariance structure changes by tracking the projection energy of observations onto an estimated signal subspace.

\item We characterize the leading-order immediate-change expected detection delay (EDD) of the proposed procedure under an average run length constraint and derive closed-form choices of the window length and drift parameter that minimize this asymptotic approximation.

\item We establish an oracle-relative asymptotic efficiency result, with an explicit efficiency constant that quantifies the loss caused by heterogeneous spike strengths.

\item We develop a practical parallel procedure for the unknown-rank setting and evaluate the method through synthetic experiments and robotic-swarm case studies.
\end{itemize}

Compared with prior work, the main extension from the rank-one spiked-covariance
setting studied in \cite{xie2020sequential} to the current work is that we allow
the signal subspace to have dimension \(d>1\), which is important for applications
such as the swarm-monitoring examples considered here. We build on the general idea
of estimating signal directions from sliding windows and using the resulting projection
energy for sequential detection, but extend both the statistic and the analysis to a
multi-dimensional signal subspace. We also introduce a separate parallel implementation
for the practical case where the rank is unknown. The theoretical results are developed
for an arbitrary fixed rank \(d\) under the stated assumptions.

\subsection{Related Work}

Early approaches to change-point detection treated covariance shifts within the classical statistical process control. Traditional Shewhart control charts and CUSUM procedures \citep{page1954continuous} were extended to monitor changes in covariance in low-dimensional settings. For instance, Hotelling's $T^2$ control chart \citep{hotelling1947multivariate} can detect general mean or covariance shifts, and methods based on determinants or generalized likelihood have been used to detect known covariance changes. Also, \cite{healy1987note} studied a multivariate CUSUM procedure for a covariance shift scaled by a constant. Similarly, \cite{edward1985multivariate} proposed using the determinant of the sample covariance as a univariate statistic to flag a covariance change. However, these classical techniques often assumed either that the post-change covariance was fully specified or that the number of data dimensions was modest.

More recent works have addressed covariance change detection in high-dimensional settings with unknown change-points in an offline manner. Notably, \cite{chen2004statistical} developed hypothesis tests for a change in a Gaussian covariance, using information criteria to estimate the change-point. \cite{wang2021optimal} proposed a binary segmentation approach to localize multiple covariance breaks, while \cite{avanesov2018change} introduced a test based on differences in inverse covariance matrices. \cite{arias2012detection} considered likelihood ratio tests for the appearance of a correlation between a specific subset of variables.

Despite these advances in offline analysis, the problem of sequentially detecting changes in covariance structure in real time remains relatively underexplored. \cite{jiao2018subspace} studied the sequential detection of a subspace switch using a CUSUM-type procedure; however, their method requires knowing the post-change subspace in advance. Recently, \cite{dornemann2026detecting} took a random matrix theory perspective to devise a hypothesis test for changes in the leading eigenvalues of a high-dimensional covariance sequence. Our work builds most directly on \cite{xie2018first, xie2020sequential}, who introduced sequential procedures for detecting when the covariance matrix shifts to a spiked model with an unknown principal component. In particular, they proposed a sliding-window chart of the largest eigenvalue and a subspace-tracking CUSUM-type detector for a single-spike model.

Moreover, numerous methods have been developed to detect structural changes in evolving networks. \cite{zhang2023spectral} proposed a spectral CUSUM method to monitor changes in community structure by inspecting the eigen-decomposition of graph-related matrices. \cite{peel2015detecting} introduced a technique to flag shifts in the large-scale connectivity patterns of an evolving network, and \cite{wang2017fast} proposed a multi-step algorithm for detecting such shifts in dynamic social networks. \cite{marangoni2014sequential} developed a sequential approach based on a spectral CUSUM statistic for detecting the emergence of communities in networks. \cite{sulem2024graph} designed a data-driven Siamese graph neural network, enabling online detection of distributional changes in a network without prior model knowledge. \cite{wang2026multilayer} proposed an online tensor-based detection algorithm to identify changes in the latent connectivity structure across layers. Notably, \cite{chen2023discovering} introduced a spectral iso-mirror embedding to reduce a dynamic network series to a scalar representation, and \cite{chen2024euclidean} demonstrated that a spectral Euclidean mirror technique can effectively localize ``first-order'' change-points in network evolution.

Our work is closely related to recent modeling efforts on swarm dynamics, which provide a foundation for the datasets considered here. In theoretical swarm dynamics, introducing range-dependent communication delays has been shown to induce rich bifurcation structures, such as torus bifurcations that transform periodic rotational swarming into quasi-periodic collective motion \citep{schwartz2020torus}. Likewise, colliding swarms exhibit a critical coupling threshold beyond which they form a single coherent milling state instead of scattering, corresponding to a saddle-node bifurcation in the swarm-on-swarm dynamics \citep{hindes2021critical}. Extending these foundations, \cite{kamimoto2023chaotic} demonstrated that when two swarms collide and merge into a combined milling state, their collective motion can transition from periodic to chaotic as key interaction parameters are varied.

\section{PROBLEM SETUP}\label{sec:problem_setup}

In this section, we formalize two closely related formulations. In the \emph{emerging subspace} setting, the data arrive as pure noise until an unknown time $\tau$ when a low-dimensional spike is added to the covariance. In the \emph{switching subspace} problem, the nominal spike shifts to an unknown direction at time $\tau$. Both problems are naturally modeled by a spiked covariance model \citep{johnstone2001distribution}.

In the \emph{emerging subspace} formulation, the observations $x_t \in \Real^k$ are independent, following a Gaussian distribution with identity covariance prior to the change, and a spiked covariance structure after the change:
\begin{align}
\begin{split}\label{eq:emerging_subspace}
&x_t \stackrel{\text{i.i.d.}}{\sim} N(0, \sigma^2 I_k), \quad t = 1, \dots, \tau, \\
&x_t \stackrel{\text{i.i.d.}}{\sim} N(0, \sigma^2 I_k + U \Lambda U^T), \quad t = \tau+1, \tau+2, \dots,
\end{split}
\end{align}
where $\tau$ is the unknown change-point. The semi-orthogonal matrix $U \in \mathbb{R}^{k \times d}$ contains $d$ orthonormal columns representing the directions of the emerging signal subspace, so that $U^TU = I_d$, and $\Lambda = \operatorname{diag}(\lambda_1, \dots, \lambda_d) \in \Real^{d \times d}$ is a diagonal matrix of signal strengths, with $\lambda_1 \ge \cdots \ge \lambda_d > 0$. The scalar $\sigma^2 > 0$ denotes the noise variance, which is assumed known. Note that in many process-monitoring applications, baseline noise levels are estimated from Phase-I or calibration data before online monitoring begins \citep{montgomery2020introduction}; assuming $\sigma^2$ known is also a theoretical simplification. We define the componentwise signal-to-noise ratio (SNR) as $\rho_i = \lambda_i / \sigma^2$ for $i = 1,\dots,d$.

Alternatively, in the \emph{switching subspace} model, both the pre- and post-change covariances have low-rank spikes, but with possibly different subspaces:
\begin{align}
\begin{split}\label{eq:switching_subspace}
&x_t \stackrel{\text{i.i.d.}}{\sim} N(0, \sigma^2 I_k + U_1 \Lambda U_1^T), \quad t = 1, \dots, \tau, \\
&x_t \stackrel{\text{i.i.d.}}{\sim} N(0, \sigma^2 I_k + U_2 \Lambda U_2^T), \quad t = \tau+1, \tau+2, \dots,
\end{split}
\end{align}
Here, $U_1, U_2 \in \Real^{k \times d}$ are both semi-orthogonal matrices representing the pre- and post-change signal subspaces, satisfying $U_1^T U_1 = U_2^T U_2 = I_d$. In this setting, the baseline subspace $U_1$ is assumed to be known \textit{a priori} from the nominal distribution.

It is worth noting that the switching subspace model can be reduced to the emerging subspace formulation by projecting the data onto the orthogonal complement of $U_1$. Specifically, we can take a matrix $Q \in \mathbb{R}^{(k-d) \times k}$ that satisfies $Q U_1 = 0$ and $Q Q^T = I_{k-d}$, and define the projected observations
\[
    y_t = Q x_t \in \Real^{k-d}, \quad t=1,2,\dots.
\]
Then, each $y_t$ is Gaussian with pre-change covariance $\sigma^2 I_{k-d}$ and
post-change covariance $\sigma^2 I_{k-d} + (Q U_2)\Lambda(Q U_2)^T.$
Here, the post-change covariance matrix
has rank at most $d$
and admits an eigendecomposition
\[
(Q U_2)\Lambda(Q U_2)^T
= U_y \Lambda_y U_y^T,
\]
where $U_y \in \Real^{(k-d)\times r}$ has orthonormal columns, $\Lambda_y=\operatorname{diag}(\lambda_{y,1},\ldots,\lambda_{y,r})$ with
$\lambda_{y,j}>0$, and $r\le d$ is the effective post-change rank.
In the case where $\operatorname{col}(U_2)\cap\operatorname{col}(U_1)=\{0\}$, we have  $r=d$;
otherwise, components of $U_2$ lying in $\operatorname{col}(U_1)$ are annihilated
by the projection, and the effective rank is reduced.

Therefore, after projection, the post-change distribution can always be written in
standard spiked-covariance form as
\[
y_t \sim N(0, \sigma^2 I_{k-d} + U_y \Lambda_y U_y^T),
\]
with signal subspace $\operatorname{col}(U_y)$ and spike strengths $\lambda_{y,1},\ldots,\lambda_{y,r}$.
Thus, the switching subspace problem can be treated within the same analytical framework as the emerging subspace case.
We note that this reduction involves a loss of information due to projection onto a lower-dimensional subspace. Nonetheless, it enables a unified, computationally efficient detection strategy applicable to both emerging and switching subspace scenarios. Accordingly, throughout the rest of this paper, we restrict attention to the emerging subspace model and develop the detection procedures and theoretical analysis in that setting, noting that the switching subspace problem can be handled via the projection-based reduction described above.

\begin{remark}[Geometric interpretation via principal angles]
Let $\theta_1,\ldots,\theta_d \in [0,\pi/2]$ denote the principal angles between the
subspaces $\operatorname{col}(U_1)$ and $\operatorname{col}(U_2)$.
Then the matrix
\[
U_2^T (I - U_1 U_1^T) U_2
\]
has eigenvalues $\sin^2(\theta_1),\ldots,\sin^2(\theta_d)$.
In particular, when $\Lambda=\lambda I_d$, the nonzero eigenvalues of the projected
spike term $(Q U_2)\Lambda(Q U_2)^T$ are given by
\[
\lambda_{y,j} = \lambda \sin^2(\theta_j), \qquad j=1,\ldots,d,
\]
with those equal to zero corresponding to $\theta_j=0$, i.e., directions in $\operatorname{col}(U_2)$ that lie in $\operatorname{col}(U_1)$ and are annihilated by the projection.
More generally, for diagonal $\Lambda=\operatorname{diag}(\lambda_1,\ldots,\lambda_d)$,
the projected spike strengths are the eigenvalues of
\[
\Lambda^{1/2} U_2^T (I - U_1 U_1^T) U_2\Lambda^{1/2},
\]
which couples the original spike magnitudes with the relative orientation of the two subspaces.
\end{remark}

A sequential detection procedure is represented by a stopping time \(T\) with respect
to the filtration \(\{\Fc_t\}\), where \(\Fc_t=\sigma(x_1,\ldots,x_t)\). We use
\(\P_{\infty}\) and \(\E_{\infty}\) to denote probability and expectation when no
change occurs, and \(\P_{\tau}\) and \(\E_{\tau}\) when the change occurs at time
\(\tau\). The average run length (ARL) to false alarm is defined as
\(
\E_{\infty}[T].
\)
For a prescribed ARL level \(\gamma>1\), we consider procedures satisfying
\(
\E_{\infty}[T]\ge \gamma.
\)
In the theoretical analysis below, we focus on the immediate-change expected detection
delay (EDD),
\(
\E_0[T],
\)
where the post-change regime starts at the beginning of the monitoring horizon.
\begin{remark}\label{rem:worst_case_delay}
A commonly used worst-case performance metric is Lorden's delay criterion
\citep{lorden1971procedures},
\begin{equation}
    \sup_{\tau\ge0}
    \operatorname*{ess\,sup}
    \mathbb E_\tau\!\left[(T-\tau)^+\mid \mathcal F_\tau,\; T>\tau\right].
    \label{WEDD}
\end{equation}
Using Lemma~4 in \citet{xie2023window}, one can show that for the
window-limited adaptive CUSUM procedure, the immediate-change delay \(\E_0[T]\)
provides an upper bound for \eqref{WEDD}. Our proposed procedure is a special case
of this window-limited adaptive CUSUM construction. Therefore, in the theoretical
analysis below, we characterize \(\E_0[T^{\rm sub}]\).
\end{remark}

\subsection{Motivating Application: Robotic Swarm Monitoring}

The emerging and switching subspace problems are directly relevant to modeling and analyzing collective dynamics in swarms. For example, consider a scenario in which a group of autonomous agents moves cohesively, forming stable formations characterized by low-dimensional subspaces in their positional and velocity data. When the swarm undergoes structural transitions such as splitting into subgroups, merging, or forming new coherent structures, the underlying covariance structure of the agents' movements changes accordingly, resulting in the emergence or shift of principal subspaces.

The rationale for modeling structural changes via low-rank covariance shifts is grounded in the intrinsic dimensionality commonly observed in real-world data. In many high-dimensional applications, including image sequences, network behaviors, and swarm trajectories, temporal dynamics typically concentrate energy in a few dominant directions, with the remaining dimensions largely capturing noise or insignificant variation. As illustrated in \cite[Figure 2]{xie2016sequential}, practical datasets often exhibit sharp transitions in spectral characteristics, where a small number of principal components abruptly emerge or vanish at a change-point. Leveraging a low-rank representation not only captures these fundamental structural patterns but also enhances sensitivity to subtle yet meaningful variations, thereby enabling quicker and more reliable change-point detection.

\section{MULTI-RANK SUBSPACE-CUSUM}\label{sec:method}

Before presenting our Multi-rank Subspace-CUSUM (MRS-C), we briefly recall the classical CUSUM test proposed by \cite{page1954continuous}. When the pre-change and post-change densities are both known, this Exact CUSUM test is optimal for sequential change detection \citep{moustakides1986optimal}. In our setting, this would require full knowledge of the post-change signal subspace $U$ and the spike magnitudes $\lambda_i$, quantities that are unavailable in practice. The Exact CUSUM is therefore an oracle benchmark. MRS-C, discussed in Section~\ref{subsec:subspace_cusum}, estimates the signal subspace on the fly, enabling a practical procedure that approximates this oracle benchmark procedure.

\subsection{Oracle Procedure: Exact CUSUM}\label{subsec:exact_cusum}

The (Exact) CUSUM test \citep{page1954continuous} uses the log-likelihood ratio between the post-change and pre-change densities. The CUSUM statistic is defined as the maximum of the accumulated log-likelihood ratio over all possible change-point locations up to time $t$:
\[
S_t = \max_{1 \le j \le t} \sum_{i=j}^{t} \log \frac{f_{0}(x_i)}{f_{\infty}(x_i)},
\]
where $f_{\infty}(\cdot)$ and $f_{0}(\cdot)$ denote the probability density functions of an observation before and after the change, respectively.
Equivalently, $S_t$ can be computed recursively as \citep{moustakides1986optimal}
\[
S_t = (S_{t-1})^+ + \log \frac{f_{0}(x_t)}{f_{\infty}(x_t)}, \quad S_0=0,
\]
where $x^+ = \max\{0, x\}$. That is, at each time we add the new log-likelihood ratio and reset it to zero if the sum becomes negative. The stopping time for the CUSUM procedure is the first time $t$ that $S_t$ exceeds a preset threshold $b$ chosen to control the false alarm rate:
\[
T_C = \inf\{t>0: S_t \ge b\}.
\]
Under the emerging subspace model \eqref{eq:emerging_subspace} that we study in this paper, the log-likelihood ratio for an observation $x_t$ is obtained as
\begin{align*}
    \log \frac{f_0(x_t)}{f_\infty(x_t)}
    &= \log \left[ \frac{\left[ (2\pi)^k \abs{\sigma^2 I_k + U \Lambda U^T} \right]^{-1/2} \exp \left\{ -\frac{1}{2} x_t^T \left(\sigma^2 I_k + U \Lambda U^T\right)^{-1} x_t \right\}}{\left[ (2 \pi)^k \sigma^{2k} \right]^{-1/2} \exp\left\{ -\frac{1}{2 \sigma^2} x_t^T x_t \right\}} \right] \\
    &=\frac{1}{2 \sigma^2}\sum_{i=1}^d \left[ \frac{\rho_i}{1 + \rho_i} (u_i^T x_t)^2 - \sigma^2  \log(1 + \rho_i) \right],
\end{align*}
The Exact CUSUM test statistic can thus be implemented by updating for each new sample $x_t$ as
\begin{align}
    S_t = (S_{t-1})^+ + \sum_{i=1}^d \left[ \frac{\rho_i}{1 + \rho_i} (u_i^T x_t)^2 - \sigma^2 \log(1+\rho_i) \right],
\end{align}
omitting the multiplicative factor
$(2 \sigma^2)^{-1} > 0$. 
When $S_t$ exceeds the threshold $b$, we declare that a change has been detected.

Notably, the update term exhibits negative drift before the change and positive drift after the change. Indeed, letting $\E_\infty[\cdot]$ denote expectation under the pre-change distribution and $\E_0[\cdot]$ the expectation under the post-change distribution, one can see that
\begin{align}\label{eq:notKL_infty}
    \E_{\infty}\left[\log\frac{f_0(x_t)}{f_\infty(x_t)}\right] = -\frac{1}{2}\sum_{i=1}^d \Big[ \log(1+\rho_i) - \frac{\rho_i}{1+\rho_i} \Big] < 0,
\end{align}
while under the post-change regime,
\begin{align}\label{eq:KL_zero}
\E_{0}\left[\log\frac{f_0(x_t)}{f_\infty(x_t)}\right] = \frac{1}{2}\sum_{i=1}^d \left[\rho_i - \log(1+\rho_i)\right] > 0.
\end{align}
This ensures $S_t$ will tend to drift downwards and remain close to zero in the absence of change, but will drift upwards after the change, as desired.

\subsection{MRS-C Procedure}\label{subsec:subspace_cusum}

In practice, the Exact CUSUM requires \emph{a priori} knowledge of the post-change subspace $U$ and signal strengths ${\lambda_i}$, which is often unavailable. A straightforward generalized likelihood ratio approach
would involve continuously re-estimating these parameters and plugging them into the likelihood ratio, but doing so naively at every time $t$ would be computationally intensive. Instead, we propose the Multi-rank Subspace-CUSUM (MRS-C) procedure, which tracks an estimated signal subspace in real time and uses it to construct a scalable detection statistic.

We maintain an estimate $\hat{U}_t \in \mathbb{R}^{k\times d}$ of the leading $d$-dimensional signal subspace at time $t$. At each time step, this estimate is obtained by forming the sample covariance matrix over a sliding window of length $w$,
\begin{align}\label{eq:cov_matrix}
\hat{\Sigma}_t
= \frac{1}{w} \sum_{i=t-w+1}^t x_i x_i^T,
\end{align}
and extracting its $d$ leading eigenvectors.
Specifically, we let $\hat{U}_t = [\hat u_{1,t},\dots,\hat u_{d,t}]$ denote the matrix of unit-norm eigenvectors of $\hat{\Sigma}_t$ corresponding to its $d$ largest eigenvalues. Under the post-change regime, $\hat{U}_t$ provides a plug-in estimate of the true signal subspace $U$.

The window length $w$ controls the trade-off between estimation accuracy and adaptivity. Larger values of $w$ improve subspace estimation but delay responsiveness to changes in the covariance structure.
Before the change, the estimate reflects only noise fluctuations, whereas after the appearance of a low-rank spike, it progressively aligns with $U$.
Accordingly, the choice of $w$ directly affects the detection delay and must be balanced against the desired false-alarm performance.

When estimating the signal subspace at each time step, we deliberately use a delayed update. Specifically, the score associated with $x_t$ is computed at time $t+w$, using the window $\{x_{t+1},\ldots,x_{t+w}\}$ to estimate the signal subspace. The key advantage of this delayed-window construction is that the subspace estimate is independent of the scored observation \(x_t\), which greatly simplifies the drift calculations and theoretical analysis presented in Section \ref{sec:analysis}.

Given such a delayed subspace estimate $\hat{U}_{t+w}$ computed over the time interval $[t+1,t+w]$, we quantify how ``signal-like'' the observation $x_t$ is through a scalar increment $Z_t$.
A natural choice is the squared norm of the projection of $x_t$ onto the estimated
subspace,
\begin{align}\label{eq:increment}
    Z_t = \big\| \hat{U}_{t+w}^T x_t \big\|^2
    = \sum_{i=1}^d \big(\hat{u}_{i,t+w}^T x_t\big)^2 .
\end{align}
Thus, $Z_t$ measures the projection energy of the observation $x_t$ onto the subspace estimated from the subsequent disjoint window.

Accordingly, we define the MRS-C statistic $S_t^{\mathrm{sub}}$ recursively as
\begin{align}\label{eq:subspace_cusum}
S_t^{\mathrm{sub}} = ( S_{t-1}^{\mathrm{sub}} )^+ + Z_t - \Delta, \quad S_0^{\mathrm{sub}}=0.
\end{align}
The drift parameter $\Delta$ should be set between the typical values of $Z_t$ under the pre-change and post-change scenarios. Formally, we require the double inequality
\begin{align}\label{eq:drift_condition}
\E_{\infty}[Z_t] < \Delta < \E_{0}[Z_t],
\end{align}
so that $S_t^{\mathrm{sub}}$ has a negative drift before the change and a positive drift after the change. Selection of the drift will be discussed rigorously in Section \ref{subsec:drift}.
The detection time is
\begin{align}\label{eq:subspace_cusum_detection_time}
    T^{\mathrm{sub}} = \inf\{t: S_t^{\mathrm{sub}} \ge b\} + w,
\end{align}
for a threshold $b$ calibrated via Monte Carlo simulation to achieve the desired false alarm rate. The stopping time includes the additional ``$+w$'' term because the detection statistic indexed by $t$ is available only at time $t+w$.

The MRS-C procedure, therefore, involves a trade-off between subspace-estimation accuracy and detection speed. Before the procedure can reliably raise an alarm, the delayed estimate $\hat{U}_t$ must capture the new spike directions. This estimation window introduces an inherent delay after the true change-point during which $\hat U_t$ is converging to $U$ and $S_t^{\mathrm{sub}}$ may not yet be growing. Once $\hat{U}_t$ sufficiently approximates $U$, the statistic adds increments $Z_t - \Delta$ with positive mean, and thereafter $S_t^{\mathrm{sub}}$ behaves similarly to the Exact CUSUM.

\section{THEORETICAL ANALYSIS}\label{sec:analysis}

This section develops the theoretical performance guarantees for the proposed MRS-C procedure under the emerging subspace model. We first characterize the mean behavior of the detection statistic before and after the change, which leads to principled choices of the drift parameter and the window length. Building on these results, we derive asymptotic expressions for the EDD under a prescribed ARL constraint and compare the delay with that of the oracle Exact CUSUM.

\subsection{Drift Parameter}\label{subsec:drift}

In this subsection, we analyze the drift parameter $\Delta$ of the detection statistic under the rank-$d$ spiked covariance model. Recall that the statistic is updated at each time step by adding the quantity $Z_t - \Delta$, with a reset to zero whenever the cumulative sum becomes negative. 
To ensure correct operation of the procedure, $\Delta$ must lie  between the pre-change and post-change expectations of $Z_t$. Accordingly, our  first task is to characterize $\E_\infty[Z_t]$ and $\E_0[Z_t]$  under the two regimes.

Importantly, by our procedure design, the subspace estimate $\hat{U}_{t+w}$ is computed from a moving window that does not include the observation $x_t$. This ensures the independence of $\hat{U}_{t+w}$ and $x_t$, which allows us to simplify the expectation as
\begin{align*}
    \E Z_t = \E \big[ \E[Z_t \big| \hat U_{t+w}] \big] = \E \Big[ \tr \big[\hat U_{t+w}^T \E[x_t x_t^T] \hat U_{t+w}\big]\Big].
\end{align*}

\paragraph*{Pre-change increment.} Before the change, $x_t$ follows $N(0, \sigma^2 I_k)$ with no signal component. Since $\E[x_t x_t^T] = \sigma^2 I_k$ and $\hat U_{t+w}$ is semi-orthogonal, we have exactly
\begin{align}
\label{eq:pre_change_inc}
    \E_\infty Z_t = d\sigma^2.
\end{align}

\paragraph*{Post-change increment.} To specify the post-change increment, we utilize the asymptotic properties of the sample eigenvectors. Throughout Section \ref{sec:analysis}, we impose the following assumption for theoretical analysis.
\begin{assumption}\label{assumption:covariance_structure}
    Assume the following for the true subspace covariance structure.
    \begin{enumerate}
        \item[(i)] The subspace dimension $d$ remains a constant.
        \item[(ii)] The population eigenvalues $\{\lambda_i: i = 1, \dots, d\}$ are distinct and are ordered as $\lambda_1 > \dots > \lambda_d > 0$.
    \end{enumerate}
\end{assumption}
Under Assumption \ref{assumption:covariance_structure}, each sample eigenvector $\hat{u}_{i, t+w}$ is approximately Gaussian distributed around the true eigenvector $u_i$, with covariance scaling as $w^{-1}$ \citep{anderson1963asymptotic, paul2007asymptotics}. Based on this approximation, we derive the following expression for the post-change mean of the increment. A detailed derivation can be found in Appendix \ref{app:proofs}.

\begin{remark}[Eigenvalue multiplicity]
Assumption~\ref{assumption:covariance_structure}(ii) requires distinct population spike eigenvalues \(\{\lambda_i\}\).
This condition is imposed for technical convenience in Lemma~\ref{lem:post_change_increment}, where distinct population eigenvalues make the individual population eigenvectors identifiable and allow us to invoke standard sample-eigenvector perturbation results \citep{anderson1963asymptotic, paul2007asymptotics}. The implementation of MRS-C does not require such distinctness. The statistic depends on the estimated signal subspace through the projection matrix, rather than on the ordering or separation of individual eigenvalues. Therefore, the procedure remains well-defined when population spike eigenvalues have multiplicities, as in several of our simulation settings in Section~\ref{sec:experiments}, although extending the perturbation expansion in Lemma~\ref{lem:post_change_increment} to that case would require a more involved argument.
\end{remark}

\begin{lem}[Post-change increment]\label{lem:post_change_increment}
    Under Assumption \ref{assumption:covariance_structure}, the expected increment under the post-change regime is given by, as $w \to \infty$,
    \begin{align}\label{eq:post_change_inc}
        &\E_0 [Z_t] = \sigma^2 \sum_{i=1}^d (1+\rho_i) - \frac{(k-d)\sigma^2}{w} \sum_{i=1}^d \frac{1+\rho_i}{\rho_i}+ o\Big(\frac{1}{w}\Big).
    \end{align}
\end{lem}

Again, for the MRS-C procedure to function as intended, the drift parameter
\(\Delta\) must satisfy the condition \eqref{eq:drift_condition}.
Comparing \eqref{eq:pre_change_inc} and \eqref{eq:post_change_inc}
suggests choosing the window size \(w\) sufficiently large so that the leading
signal contribution dominates the window-induced estimation error. At the leading
order, this amounts to
\[
w >
\frac{k-d}{\sum_{i=1}^d \rho_i}
\sum_{i=1}^d \frac{1+\rho_i}{\rho_i},
\]
under which the leading-order approximation to the post-change mean exceeds
the pre-change mean. This condition is also consistent with the large-window regime
in which the Gaussian perturbation approximation for the sample eigenvectors becomes
accurate. Thus, for sufficiently large \(w\), the expansion suggests a nonempty
leading-order range for choosing the drift parameter \(\Delta\).

In practice, the exact signal strengths $\{\lambda_i\}$ may not be known in advance. However, we can design $\Delta$ based on a conservative lower bound for the signal strength.  For example, suppose it is known that each signal component has SNR at least $\rho_{\min} = \lambda_{\min}/\sigma^2$. A robust choice is to set $\Delta$ to the midpoint between the pre-change expectation and a conservative estimate of the post-change expectation as
\[
\Delta = d\sigma^2 + \frac{1}{2}d\lambda_{\min} = d\sigma^2\Big(1 + \frac{1}{2}\rho_{\min}\Big).
\]
This places \(\Delta\) above the pre-change mean \(d\sigma^2\), while keeping the drift choice conservative when the spike strength is uncertain. The empirical sensitivity of MRS-C to the design value of \(\rho_{\min}\) is examined in Appendix~\ref{app:rho_sensitivity}.

\begin{remark}
    Equation \eqref{eq:post_change_inc} shows that the leading term in the post-change expectation depends only on the noise level $\sigma^2$, the signal-to-noise ratios $\rho_i$, and the window size $w$. Importantly, this term is independent of the specific orientation of the signal subspace $U$. Therefore, to estimate $\E_0 [ \lVert \hat U^T x_t \rVert^2 ]$ in practice, one may simulate the Gaussian samples using a randomly generated semi-orthogonal matrix $U_0 \in \mathbb{R}^{k \times d}$ and covariance $\sigma^2 I_k + U_0 \Lambda U_0^T$.
\end{remark}

\subsection{Asymptotic EDD and Oracle-relative Efficiency}\label{subsec:asymp_opt}
In this section, we study the asymptotic performance of the proposed MRS-C procedure. Building on the drift analysis in Section~\ref{subsec:drift}, we derive an asymptotic expression for the EDD as the ARL tends to infinity.
We then use this result to identify asymptotically optimal choices of the threshold $b$, drift $\Delta$, and window length $w$ that minimize the asymptotic EDD, and characterize the oracle-relative efficiency of MRS-C compared with the Exact CUSUM procedure.

The following Theorem~\ref{thm:arl-edd} provides an explicit asymptotic
characterization of the EDD for the MRS-C procedure. We first introduce the following notion of local dependence.
\begin{defn}[$w$-dependence \citep{janson1984runs}]\label{def:w_dependence}
For an integer $w\ge 0$, a discrete-time stochastic process
$\{Y_t\}_{t\ge 1}$ is called $w$-dependent if, for every $k\ge 1$,
the sigma-fields
\[
\sigma(Y_1,\ldots,Y_k)
\quad\text{and}\quad
\sigma(Y_{k+w+1},Y_{k+w+2},\ldots)
\]
are independent. Equivalently, any two collections of variables separated by more
than $w$ time indices are independent.
\end{defn}
By construction, $Z_t - \Delta$ depends only on the raw-data block
$\{x_t,x_{t+1},\ldots,x_{t+w}\}$. Therefore, $\{Z_t - \Delta\}$ are $w$-dependent,
placing our analysis in the same dependence
setting as window-limited CUSUM procedures \citep{xie2023window}.

\begin{thm}[Asymptotic EDD of MRS-C]\label{thm:arl-edd}
For a prescribed ARL level $\E_\infty[T^{\mathrm{sub}}] = \gamma > 1$, suppose as $\gamma \to \infty$, $w \asymp \sqrt{\log \gamma}$. If we select the threshold $b$ and the drift $\Delta$ as
\begin{align}
    \label{eq:threshold_and_A}b &= \frac{2 \sigma^2 \log \gamma(1 + o(1))}{1 - d/A},\\
    \label{eq:drift_and_A}\Delta &= \frac{d \sigma^2}{1 - d/A} \log\Big(\frac{A}{d}\Big),
\end{align}
where $A = A(w)$ represents the dominant term of the post-change increment expectation as
\begin{align}\label{eq:term_A}
    A(w) \coloneqq \sum_{i=1}^d (1 + \rho_i) \Big(1 - \frac{k-d}{w \rho_i}\Big),
\end{align}
then the EDD under the alternative is given by
\begin{align}\label{eq:edd}
    \E_0[T^{\mathrm{sub}}] = \frac{2 \log \gamma (1 + o(1))}{A -d (1 + \log (A/d))}  + w.
\end{align}
\end{thm}

Theorem~\ref{thm:arl-edd} is stated for the immediate-change EDD. The delayed-window construction considered here is a special case of the window-limited adaptive CUSUM in \citep{xie2023window}. Indeed, following the argument of Lemma~4 in
\citep{xie2023window}, \(E_0[T^{\mathrm{sub}}]\) provides an upper bound for the
Lorden-type worst-case delay in this setting.

In particular, Theorem~\ref{thm:arl-edd} identifies a pair $(b,\Delta)$ through the pre-change adjustment-coefficient approximation. 
The condition
\[
\E_\infty\left[e^{\delta_\infty(Z_t-\Delta)}\right]=1
\]
leads to a martingale construction that enables us to find the EDD approximation \eqref{eq:edd}. The detailed proof is given in Appendix \ref{app:proofs}.

The expression for $A(w)$ in \eqref{eq:term_A} represents the leading-order post-change drift of the statistic, incorporating both the spike strengths $\{\rho_i\}$ and the window-induced eigenvector estimation error at order  $w^{-1}$. The EDD expression in \eqref{eq:edd} is decomposed into two components: the intrinsic information-theoretic delay, and the estimation lag $w$ due to the sliding window needed to estimate the evolving subspace independently. The intrinsic delay scales logarithmically with the ARL, as in the classical CUSUM theory.

Minimizing the leading-order EDD in Theorem \ref{thm:arl-edd} with respect to the window length $w$ yields the optimal parameters. Let
\(
    \bar\rho = \sum_{i=1}^d \rho_i / d,
\)
representing the average signal-to-noise ratio.
\begin{cor}[Optimal parameters]\label{thm:opt_window}
    To minimize leading-order terms in \eqref{eq:edd}, the asymptotically optimal window size $w^*$ is given by
    \begin{align}
        w^* = \frac{\sqrt{\log \gamma}}{\bar\rho - \log(1+\bar\rho)} \cdot\left[2\left(\frac{k}{d}-1\right)\left(1 + \overline{\rho^{-1}}\right)\left(\frac{\bar\rho}{1+\bar\rho}\right)\right]^{1/2} (1 + o(1)),
    \end{align}
    where $\overline{\rho^{-1}} = \sum_{i=1}^d \rho_i^{-1} / d$. Hence, corresponding optimal choices of the threshold $b$ and the drift $\Delta$ are given by
    \begin{align}
        b \asymp \frac{2 \sigma^2 \log \gamma}{1 - d/A(w^*)}, \quad
        \Delta = \frac{d \sigma^2}{1 - d/A(w^*)} \log\Big(\frac{A(w^*)}{d}\Big).
    \end{align}
\end{cor}

Finally, we establish the efficiency of our procedure relative to the oracle Exact CUSUM.

\begin{thm}[Oracle-relative asymptotic efficiency]\label{thm:optimality}
    The ratio of the expected detection delays of MRS-C ($T^{\mathrm{sub}}$) to the oracle Exact CUSUM ($T_C$) is given by 
    \begin{align}
        \frac{\E_0[T^{\mathrm{sub}}]}{\E_0[T_C]} = \Kc + O\Big(\frac{1}{\sqrt{\log \gamma}}\Big),
    \end{align}
    where the efficiency constant $\Kc \ge 1$ is defined as
    \begin{align}
        \mathcal{K} = \frac{ \bar\rho - \overline{\log(1+\rho)}}{\bar\rho - \log(1 + \bar{\rho})} \geq 1,
    \end{align}
    where $\overline{\log(1+\rho)} = (1/d)\sum_{i=1}^d \log(1 + \rho_i)$.
    The constant $\Kc$ attains its minimum value of 1 if and only if all signal strengths are uniform (i.e., $\rho_i = \bar \rho$ for all $i$).
\end{thm}

The constant $\Kc$ quantifies the effect of unequal spike strengths. When all spikes have the same strength, $\Kc=1$, and the asymptotic delay matches the oracle benchmark. When the spike strengths are heterogeneous, $K>1$, giving a fixed multiplicative loss relative to the oracle.

\begin{remark}[Difference from rank-one case]
Our theoretical results extend prior work on subspace-based change detection beyond the rank-one spiked covariance model. In particular, \cite{xie2020sequential,xie2018first} address the rank-one setting. Our multi-rank framework encompasses this case as a special case: when \(d=1\), the efficiency constant \(\Kc\) becomes \(1\), recovering the oracle-relative first-order asymptotic efficiency in the rank-one setting. However, for \(d>1\), Theorem~\ref{thm:optimality} shows that heterogeneity in signal strengths \(\{\rho_i\}\) introduces a penalty factor \(\Kc>1\).
Compared with \cite{zhang2023spectral}, which studies a different inverse-covariance/network setting based on adjacency matrices, our model leads to different drift terms and a different efficiency constant. Nevertheless, both settings exhibit the same optimal window scaling \(w^* \asymp \sqrt{\log \gamma}\), suggesting a common scaling behavior for window-limited CUSUM procedures involving subspace estimation.
\end{remark}

\section{PARALLEL PROCEDURE FOR DIMENSION SELECTION}\label{sec:parallel}

Throughout our discussion and analyses thus far, we have assumed that the dimension $d$ of the underlying subspace is known. In practice, however, the true subspace dimension is typically unknown and must be estimated from data or chosen adaptively. To handle the unknown $d$, we now consider running multiple detection procedures in parallel across a set of candidate dimensions $\Dc = \{ d_1, \dots, d_m\}$. Motivated by \cite{xie2013sequential}, by setting detection thresholds for each dimension individually and ensuring the overall false-alarm rate is properly controlled, the earliest detection from these parallel runs can be taken as the final decision.

Formally, for a given candidate dimension $d_k \in \Dc$, $k = 1, \dots, m$, the single-procedure update is
\[
Z_t^{(d_k)} = \sum_{i=1}^{d_k} \big(\hat{u}_{i, t+w}^T x_t\big)^2, \quad
S_t^{(d_k)} = (S_{t-1}^{(d_k)})^+ + Z_t^{(d_k)} - d_k \Delta_1,
\]
with $S_0^{(d_k)} = 0$.
Each $S_t^{(d_k)}$ is compared against a threshold $b^{(d_k)}$, where each threshold $b^{(d_k)} > 0$ is chosen to achieve the desired ARL under the window length $w^{(d_k)}$.
We declare that the procedure associated with dimension $d_k$ signals a change at time
\[
T^{(d_k)} = \inf \{t: S_t^{(d_k)} \ge b^{(d_k)}\} + w^{(d_k)}.
\]
The parallel stopping rule for the overall procedure is the minimum of these stopping times over all candidate dimensions
\begin{equation}\label{eq:Tmix}
    T_{\text{mix}} = \min_{d_k \in \Dc} T^{(d_k)} =  \min_{k \in [m]} \left\{ \inf \{ t : S_t^{(d_k)} \ge b^{(d_k)} \} + w^{(d_k)} \right\}.
\end{equation}
In words, $T_{\text{mix}}$ is the first time any one of the $m$ parallel statistics signals a change. The index achieving the minimum in \eqref{eq:Tmix}, $k^* = \argmin_k T^{(d_k)}$, effectively suggests an estimate of the post-change subspace dimension as
\begin{align}\label{eq:dhat}
\hat d = d_{k^*}.
\end{align}
The rationale is that whichever candidate dimension $d_k$ best matches the true subspace will accumulate evidence the fastest.

Under the fixed threshold regime, given the target ARL $\gamma$, we employ a simple and conservative Bonferroni-type union bound using the fact that
\[
\P_\infty(T_{\mathrm{mix}} \le t) \le \sum_{k=1}^m \P_\infty (T^{(d_k)} \le t).
\]
Therefore, each threshold $b^{(d_k)}$ is calibrated to have ARL $m \cdot \gamma$ by Monte Carlo simulation.

\begin{remark}
The fixed-rank analysis in Section~\ref{sec:analysis} can be used as a design guideline for a specified candidate dimension. When the rank is unknown, one may either assign a candidate-specific window \(w^{(d_k)}\) to each chart using \(d_k\) and a conservative SNR lower bound, or use a common window selected conservatively over the candidate set \(D\), as in the experiments in Section~\ref{subsec:parallel_experiment}. We do not claim that either choice is asymptotically optimal for \(T_{\mathrm{mix}}\).
\end{remark}

\section{NUMERICAL EXPERIMENTS}\label{sec:experiments}

We evaluate MRS-C through synthetic experiments and swarm-monitoring examples. The synthetic experiments in Sections~\ref{subsec:simulations} and~\ref{subsec:parallel_experiment} provide controlled settings where ARL calibration, EDD, and rank selection can be assessed quantitatively. The robotic-swarm examples in Section~\ref{subsec:swarm_experiment} illustrate how the statistic behaves on trajectory data where changes correspond to interpretable formation or motion transitions.

\subsection{Simulation Study}\label{subsec:simulations}

We divide the simulation study into two parts. 
Section \ref{subsubsec:simulation_finite_samples} checks finite-sample ARL calibration and compares MRS-C with the oracle Exact CUSUM and the Largest-Eigenvalue Shewhart Chart \citep{xie2020sequential}. 
Section \ref{subsubsec:simulation_asymptotic} then compares Monte Carlo ARL--EDD curves with the large-ARL approximation from Theorem \ref{thm:arl-edd}, using the window choice $w^*(\gamma)$ from Corollary \ref{thm:opt_window}.

\subsubsection{Finite-sample ARL calibration and baseline comparison}\label{subsubsec:simulation_finite_samples}

We first assess finite-sample ARL calibration and detection delay under controlled Gaussian spiked-covariance models.
We simulate sequences of $k$-dimensional  observations $x_t \sim N(0, \Sigma_t)$, where $\Sigma_t = \sigma^2 I_k$ for $t \le \tau$ before an unknown change-point $\tau$ and $\Sigma_t = \Sigma_1$ for $t > \tau$. We consider several spiked covariance scenarios for $\Sigma_1$:
\begin{itemize}
    \item \textit{Rank-1 spike}: $\Sigma_1 = \sigma^2 I_k + \lambda uu^T$, where $u$ is a unit-norm eigenvector. We consider both a dense spike where $u = (1,1,\dots,1)^T / \sqrt{k}$ and a sparse spike where $u = e_1 = (1,0,\dots,0)^T$.
    \item \textit{Rank-$d$ spike}: $\Sigma_1 = \sigma^2 I_k + U \Lambda U^T$ with a $d$-dimensional orthonormal subspace $U = [u_1,\dots,u_d]$. We consider uniform spikes $\Lambda = \lambda I_d$ where all $d$ signal eigenvalues are equal, and non-uniform spikes $\Lambda=\operatorname{diag}(\lambda_1,\dots,\lambda_d)$ with eigenvalues $\lambda_1>\lambda_2>\cdots>\lambda_d>0$ of varying strength.
\end{itemize}
In our experiments, we benchmark MRS-C against the oracle Exact CUSUM in Section~\ref{subsec:exact_cusum}, which assumes full knowledge of the post-change subspace, and the Largest-Eigenvalue Shewhart Chart (LESC) \citep{xie2020sequential}, which uses the largest sample eigenvalue of the sample covariance matrix over a sliding window. Precisely, given the rolling covariance matrix $\hat \Sigma_t$ at time $t$,
the LESC statistic is
\(
L_t =
\lambda_{\max}
(
\hat \Sigma_t
),
\)
and the corresponding stopping time is
\(
T_{\mathrm{LESC}} =
\inf
\left\{
t>0:
L_t\geq b_{\mathrm{LESC}}
\right\},
\)
where $b_{\mathrm{LESC}}$ is calibrated by Monte Carlo simulation to
attain the prescribed ARL. Unlike MRS-C, LESC is a Shewhart-type procedure that thresholds the current-window statistic without accumulating evidence over time.

Table \ref{tab:thresholds} demonstrates the MRS-C thresholds for a target ARL of 5000 under the pre-change distribution, calibrated by Monte Carlo simulation. We show results for  combinations of ambient dimension $k \in \{5, 10, 20\}$, subspace rank $d \in \{2, 3\}$, and sliding window length $w \in \{20, 50, 100\}$. In all cases, the post-change signal strength was set to $\Lambda=I_d$ with a design lower bound of componentwise SNR $\rho_{\min}=0.5$.

\begin{table}[t]
\centering
\caption{Threshold $b$ for MRS-C corresponding to target ARL of 5000, calibrated by Monte Carlo simulation: $\Lambda=I_d$, $\rho_{\min} = 0.5$. Empirical ARL values are obtained by averaging 20 independent Monte Carlo estimates, where each estimate is computed using $10^6$ renewal trials (standard errors in parentheses).}
\begin{tabular}{rcccccc}
\toprule
$w$ & \multicolumn{2}{c}{$20$} & \multicolumn{2}{c}{$50$} & \multicolumn{2}{c}{$100$} \\
\cmidrule(lr){2-3} \cmidrule(lr){4-5} \cmidrule(lr){6-7}
& $b$ & ARL & $b$ & ARL & $b$ & ARL \\
\midrule
$k=5, d=2, \sigma^2=1$ & 27.38 & 5037.2 {\scriptsize (52.5)} & 25.10 & 5006.9 {\scriptsize (56.4)} & 23.47 & 4965.5 {\scriptsize (44.1)}\\
$k=5, d=3, \sigma^2=1$ & 25.95 & 4995.2 {\scriptsize (53.7)} &24.56 & 4955.4 {\scriptsize (53.3)} & 23.35 & 5027.7 {\scriptsize (43.2)}\\
$k=10, d=2, \sigma^2=1$ & 35.18 & 4950.6 {\scriptsize (51.9)} & 30.82 & 4965.3 {\scriptsize (51.9)} & 27.24 & 4976.8 {\scriptsize (68.0)}\\
$k=10, d=3, \sigma^2=1$ & 33.12 & 5041.3 {\scriptsize (53.0)} & 29.98 & 5026.9 {\scriptsize (50.1)} & 27.18 & 4984.5 {\scriptsize (51.2)}\\
$k=20, d=2, \sigma^2=1$ & 47.82 & 4963.7 {\scriptsize (23.8)} & 40.57 & 5001.3 {\scriptsize (41.0)} & 34.61 & 5049.2 {\scriptsize (65.1)}\\
$k=20, d=3, \sigma^2=1$ & 44.84 & 5064.7 {\scriptsize (49.4)} & 39.10 & 5023.8 {\scriptsize (44.4)} & 34.03 & 4993.2 {\scriptsize (39.3)}\\
\bottomrule
\end{tabular}
\label{tab:thresholds}
\end{table}

\begin{figure}[!htb]
    \centering
    \subfigure[$\sigma^2 = 2$.]{
    \resizebox*{0.32\linewidth}{!}{\includegraphics{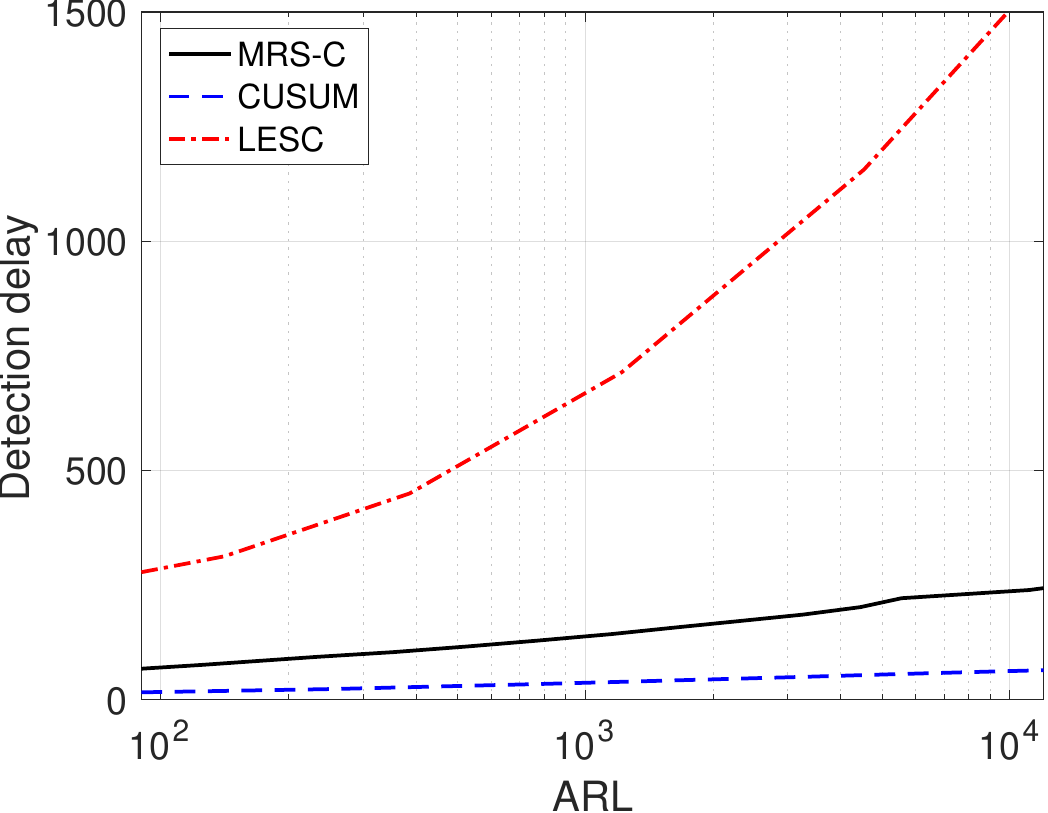}}}
    \subfigure[$\sigma^2 = 1$.]{
    \resizebox*{0.32\linewidth}{!}{\includegraphics{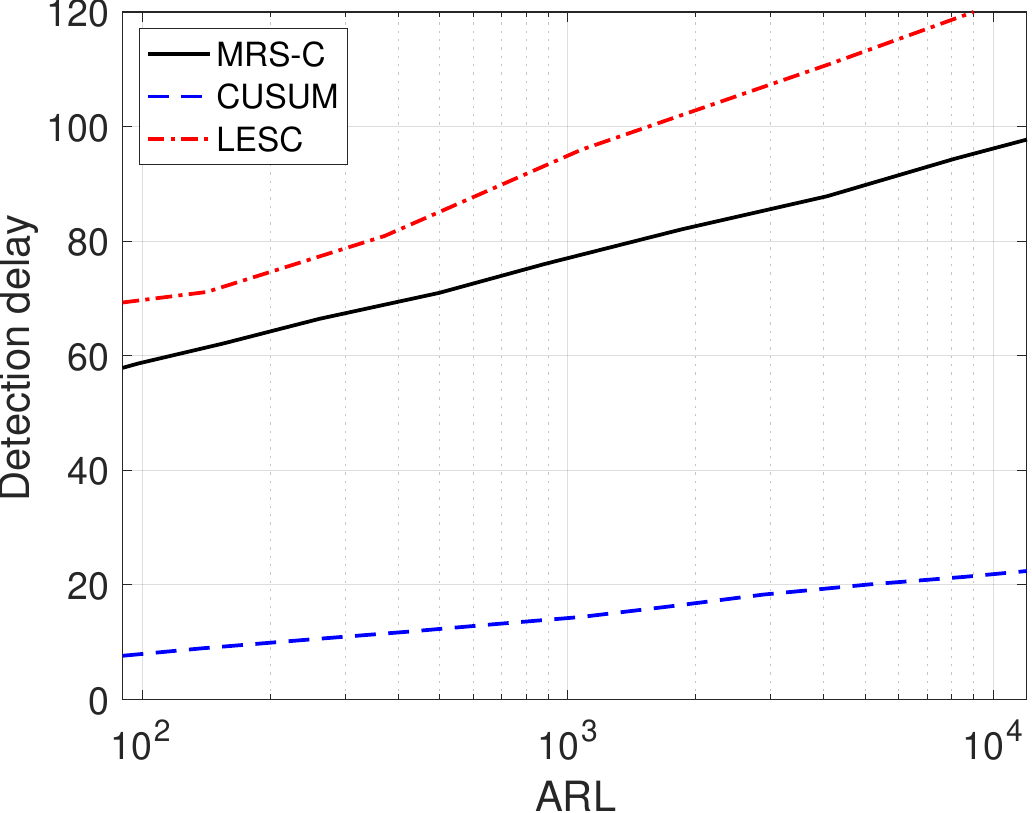}}}
    \subfigure[$\sigma^2 = 0.5$.]{
    \resizebox*{0.32\linewidth}{!}{\includegraphics{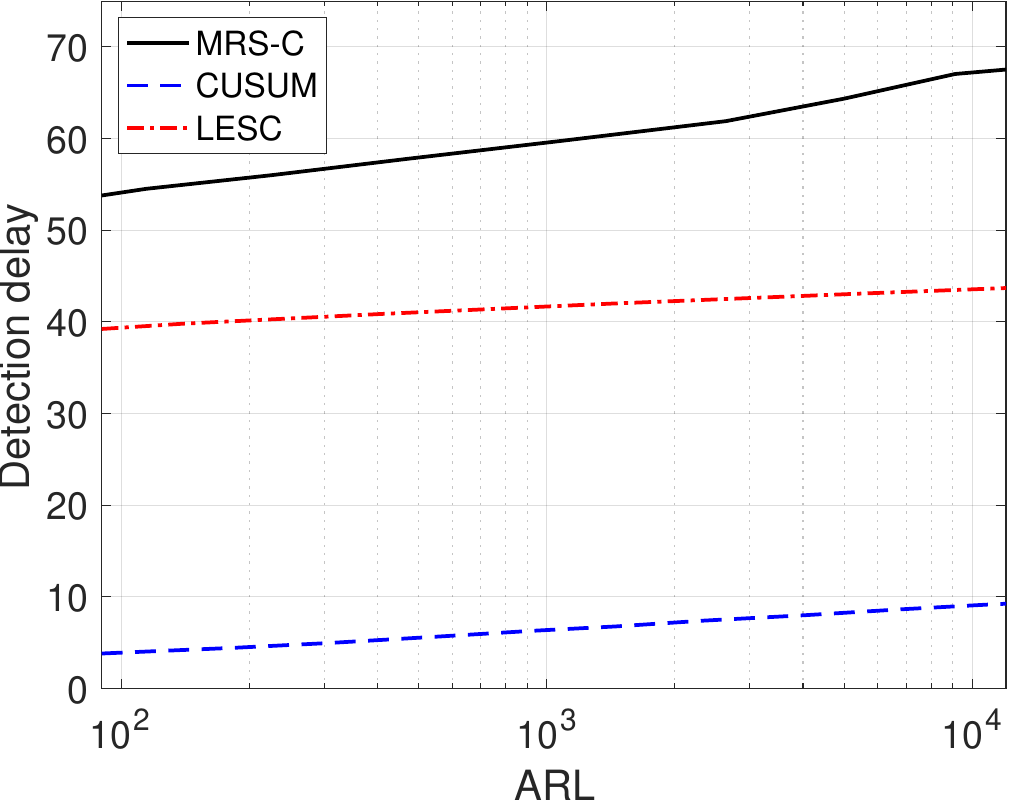}}}
    \caption{ARL versus EDD for MRS-C, the oracle Exact CUSUM, and LESC under the uniform-spike model with $k=10$, $d=2$, $\Lambda=I_2$, and varying noise levels.}
    \label{fig:ARL_EDD}
\end{figure}

Figure \ref{fig:ARL_EDD} illustrates the trade-off between ARL and EDD
under the uniform-spike model with $k=10$, $d=2$, $\Lambda=I_2$, and fixed window length $w=50$. 
The three panels correspond to different noise variances $\sigma^2 = 2, 1, 0.5$, or equivalently $\rho = 0.5, 1, 2$.
Across all noise levels, the Exact CUSUM attains the smallest detection delay for any prescribed ARL, as expected by its theoretical optimality. When the signal-to-noise ratio is low (high noise), MRS-C closely tracks the oracle performance, incurring only a modest additional delay, as MRS-C accumulates small persistent covariance evidence over time. By contrast, LESC incurs substantially larger delays under weak signals, since its one-shot decision statistic reacts slowly to small shifts. As the signal-to-noise ratio increases (low noise), LESC begins to outperform MRS-C, reflecting the well-known fact that Shewhart charts excel at detecting large, abrupt changes \citep{montgomery2020introduction}.

\begin{table}[t]
\centering
\caption{Simulated detection delays with ARL calibrated to 5000 under $\Lambda = I_d$. ``CUSUM'' and ``LESC'' denote the Exact CUSUM and the Largest-Eigenvalue Shewhart Charts, respectively (standard errors in parentheses).}
\label{tab:sim-results}
\small
\setlength{\tabcolsep}{3.5pt}
\resizebox{\textwidth}{!}{%
\begin{tabular}{cc
    *{9}{>{\centering\arraybackslash}p{0.085\textwidth}}}
\toprule
\multirow{2}{*}{$k$} & \multirow{2}{*}{$d$}
  & \multicolumn{3}{c}{$\sigma^2=2$}
  & \multicolumn{3}{c}{$\sigma^2=1$}
  & \multicolumn{3}{c}{$\sigma^2=0.5$}
\\
\cmidrule(lr){3-5} \cmidrule(lr){6-8} \cmidrule(lr){9-11}
 & 
  & CUSUM & MRS-C & LESC
  & CUSUM & MRS-C & LESC
  & CUSUM & MRS-C & LESC
\\
\midrule
5  & 2
&  52.9 \scriptsize($ 0.63$) & 159.6 \scriptsize($ 2.70$) & 850.2 \scriptsize($ 20.14$)
&  20.1 \scriptsize($ 0.12$) & 77.1  \scriptsize($ 0.76$) & 90.6  \scriptsize($ 1.67$)
&   8.4 \; \scriptsize($ 0.03$) & 63.7  \scriptsize($ 0.09$) & 40.1  \scriptsize($ 0.27$)
\\
5  & 3
&  37.9 \scriptsize($ 0.23$) & 122.3 \scriptsize($ 1.36$) & 722.8 \scriptsize($ 23.08$)
&  14.4 \scriptsize($ 0.05$) & 76.2  \scriptsize($ 0.33$) & 85.8  \scriptsize($ 1.38$)
&   6.0 \; \scriptsize($ 0.03$) & 57.8  \scriptsize($ 0.22$) & 42.9  \scriptsize($ 1.09$)
\\
10 & 2
&  52.5 \scriptsize($ 0.33$) & 196.3 \scriptsize($ 5.59$) & 1225.7 \scriptsize($ 12.30$)
&  20.2 \scriptsize($ 0.18$) & 86.8  \scriptsize($ 1.59$) & 114.1 \scriptsize($ 2.36$)
&   8.4 \; \scriptsize($ 0.05$) & 59.9  \scriptsize($ 0.41$) & 45.2  \scriptsize($ 0.36$)
\\
10 & 3
&  38.0 \scriptsize($ 0.16$) & 153.8 \scriptsize($ 2.20$) & 1043.9 \scriptsize($ 16.91$)
&  15.8 \scriptsize($ 0.10$) & 75.0  \scriptsize($ 0.44$) & 100.2 \scriptsize($ 2.76$)
&   6.0 \; \scriptsize($ 0.03$) & 62.7  \scriptsize($ 0.30$) & 50.3  \scriptsize($ 1.04$)
\\
20 & 2
&  52.4 \scriptsize($ 0.41$) & 349.1 \scriptsize($ 14.67$) & 2072.5 \scriptsize($ 43.28$)
&  20.1 \scriptsize($ 0.05$) & 106.9 \scriptsize($ 0.77$) & 185.6 \scriptsize($ 8.40$)
&   8.4 \; \scriptsize($ 0.04$) & 63.2  \scriptsize($ 0.22$) & 47.1  \scriptsize($ 0.33$)
\\
20 & 3
&  38.1 \scriptsize($ 0.28$) & 247.5 \scriptsize($ 4.05$) & 1802.0 \scriptsize($ 33.17$)
&  14.4 \scriptsize($ 0.07$) & 101.2 \scriptsize($ 1.12$) & 166.4 \scriptsize($ 7.75$)
&   6.0 \; \scriptsize($ 0.03$) & 63.2  \scriptsize($ 0.15$) & 53.7  \scriptsize($ 0.29$)
\\
\bottomrule
\end{tabular}%
}
\end{table}

Table \ref{tab:sim-results} reports detection delay values for each detector at ARL calibrated to 5000. We consider three ambient dimensions $k\in\{5,10,20\}$ and two subspace ranks $d\in\{2,3\}$, across the same three noise variances as in Figure \ref{fig:ARL_EDD}.  In every configuration, the Exact CUSUM attains the smallest delay, matching its oracle status. MRS-C remains substantially faster than LESC in low-SNR settings, while LESC becomes competitive, and sometimes faster, when the signal is strong. 
These results confirm the qualitative trade-off in Figure \ref{fig:ARL_EDD} that MRS-C is particularly useful for weak-to-moderate persistent subspace changes.

\begin{remark}[Hybrid detection strategy]\label{rem:hybrid}
The comparison with LESC suggests a natural hybrid strategy when the signal strength is uncertain. One may run the two procedures in parallel and stop at the first alarm,
\[
T_{\mathrm{hybrid}}=\min\{T^{\mathrm{sub}},T_{\mathrm{LESC}}\}.
\]
This hybrid rule provides a simple way to hedge between low- and high-SNR regimes. A fair implementation, however, requires raising the component thresholds to maintain the global ARL. We observe that the hybrid rule does not uniformly dominate the better standalone detector. In our experiments, the hybrid detector tends to be slightly slower than MRS-C in low-SNR regimes and slightly slower than LESC in high-SNR regimes; see Appendix~\ref{app:hybrid} for details. Nevertheless, it remains competitive across regimes and can be viewed as a practical alternative when the signal strength is unknown.
\end{remark}

\subsubsection{Comparison with the large-ARL approximation}
\label{subsubsec:simulation_asymptotic}

We now compare Monte Carlo ARL--EDD curves with the large-ARL approximation in Theorem~\ref{thm:arl-edd} and the window choice in Corollary~\ref{thm:opt_window}. 
The goal is to check whether the theory provides a sensible window size and captures the correct qualitative scaling as the ARL target increases.

For each ARL value \(\gamma\), we compute \(w^*(\gamma)\) from Corollary~\ref{thm:opt_window} and set the Monte Carlo window length \(w_{\mathrm{MC}}(\gamma)\) by rounding it.
We then run MRS-C with this window length and calibrate its threshold under the pre-change distribution by Monte Carlo simulation. 
Thus, unlike Section~\ref{subsubsec:simulation_finite_samples}, where window lengths are fixed, this experiment uses the same ARL-dependent window choice as the theory.

Recall that the approximate EDD from Theorem~\ref{thm:arl-edd} is
\[
    \frac{
        2\log \gamma
    }{
        A(w^*)-d\{1+\log(A(w^*)/d)\}
    }
    + w^*.
\]
This expression is derived in the large-ARL regime. 
For moderate ARL values, some finite-sample discrepancy is therefore expected.

\begin{figure}[!htb]
\centering
\includegraphics[width=\textwidth]{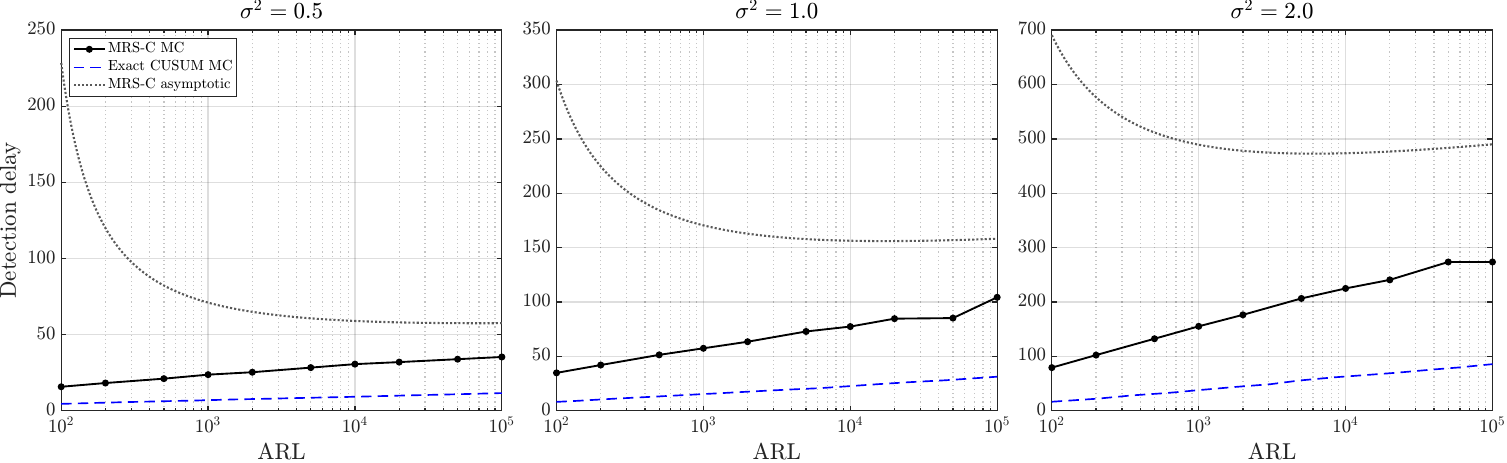}
\caption{
Monte Carlo ARL--EDD curves for MRS-C with $w_{\mathrm{MC}}=\operatorname{round}\{w^*(\gamma)\}$, compared with the large-ARL approximation from Theorem \ref{thm:arl-edd}. 
The Exact CUSUM Monte Carlo curve is included as an oracle benchmark. 
The approximation is above the Monte Carlo curve in this finite-ARL range but gives the expected dependence on signal strength.
}
\label{fig:wstar_theory_MC}
\end{figure}

Figure~\ref{fig:wstar_theory_MC} compares the \(w^*(\gamma)\)-matched MRS-C Monte Carlo curve with this approximation for \(k=10\), \(d=2\), \(\Lambda=I_2\), and \(\sigma^2=0.5,1,2\), or equivalently \(\rho=2,1,0.5\). 
The Exact CUSUM Monte Carlo curve is included as an oracle benchmark. 
MRS-C has a larger delay than Exact CUSUM, as expected, since it must estimate the unknown signal subspace. 
In these experiments, the approximation lies above the Monte Carlo curve over the displayed ARL range, especially when the signal is weak. 
The gap tends to decrease as the ARL increases, which is consistent with the fact that the formula is a large-ARL approximation.

Table~\ref{tab:wstar_theory_MC_summary} summarizes the comparison at target ARL $10^5$. 
The selected window size increases as the signal becomes weaker.
This is the expected behavior since weak signals require longer windows to estimate the signal subspace reliably.
The approximation remains larger than the Monte Carlo EDD in all three cases, but it captures the increase in delay and window length as the SNR decreases.

\begin{table}[t]
\centering
\caption{MRS-C comparison at target ARL $10^5$ for $k=10$, $d=2$, and $\Lambda=I_2$. 
For each $\sigma^2$, $w_{\mathrm{MC}}=\mathrm{round}\{w^*(10^5)\}$. Each threshold is calibrated by $10^5$ Monte Carlo renewal trials (standard errors in parentheses).}
\label{tab:wstar_theory_MC_summary}
\begin{tabular}{ccccc}
\toprule
$\sigma^2$
& $w_{\mathrm{MC}}$
& MC EDD
& Approx. EDD
& MC / Approx.\\
\midrule
0.5 & 11  & 35.4 {\scriptsize (0.05)}  & 57.6 & 0.61 \\
1.0 & 31  & 104.4 {\scriptsize (0.24)} & 158.2 & 0.66 \\
2.0 & 102 & 273.7 {\scriptsize (0.74)} & 490.0 & 0.56 \\
\bottomrule
\end{tabular}
\end{table}

This experiment clarifies the role of the theory. 
The expression in Theorem \ref{thm:arl-edd} should not be read as a precise finite-sample formula for the fixed-window curves in Section~\ref{subsubsec:simulation_finite_samples}. 
Instead, it gives a principled way to choose the window length and explains how the delay changes with the ARL and SNR. 
In finite samples, the threshold is still calibrated by Monte Carlo simulation, while $w^*(\gamma)$ provides a useful guide for choosing the window size.

\subsection{Parallel Procedure with Unknown Subspace Dimension}\label{subsec:parallel_experiment}

We next evaluate the parallel procedure proposed in Section~\ref{sec:parallel} in the setting where the post-change subspace dimension is unknown. In contrast to the experiments in Section~\ref{subsec:simulations}, here multiple MRS-C charts are run in parallel for candidate dimensions, and an alarm is raised when any one of them signals.

We consider the emerging subspace model in ambient dimension $k = 20$. 
Pre-change observations follow $N(0, I_k)$, and after the change-point fixed at $\tau = 500$, post-change observations follow
\[
x_t \sim N(0, I_k + U_{d^\star} \Lambda_{d^\star} U_{d^\star}^T),\quad 
\Lambda_{d^\star} = I_{d^\star},
\]
where $U_{d^\star} \in \mathbb{R}^{k \times d^\star}$ has $d^\star$ orthonormal columns.
We fix the window length at $w = 50$, and the candidate dimension set is $\Dc = \{1,2,\dots,10\}.$
Each individual MRS-C statistic $S_t^{(d)}$ is assigned its own threshold $b^{(d)}$, calibrated individually to achieve a per-chart ARL of 50000,
ensuring that the overall procedure across ten parallel charts achieves an ARL of 5000 via a Bonferroni-type argument.

The resulting thresholds calibrated by Monte Carlo simulations and their empirical ARLs are reported in Table~\ref{tab:parallel-10}. Table~\ref{tab:parallel-10} also summarizes rank-selection frequencies for two post-change scenarios with true ranks $d^\star = 3$ and $d^\star = 8$. In both cases, the most frequently selected dimension matches the true post-change dimension, demonstrating that the parallel procedure can infer the latent post-change dimension from the data.

\begin{table}[t!]
\centering
\caption{Calibrated per-dimension thresholds $b^{(d)}$ targeting per-chart ARL of $50000$ for each candidate dimension $d$.
Empirical ARL values are obtained by averaging 20 independent Monte Carlo estimates, where each estimate is computed using $10^5$ renewal trials (standard errors in parentheses).
The last two rows report empirical rank-selection frequencies from the parallel MRS-C procedure under two different post-change scenarios with true ranks $d^\star=3$ and $d^\star=8$, respectively.
In both cases, the mode of the selected dimension correctly matches the true rank.}
\label{tab:parallel-10}
\resizebox{\textwidth}{!}{%
\begin{tabular}{ccccccccccc}
\toprule
$d$ 
& 1 & 2 & 3 & 4 & 5 & 6 & 7 & 8 & 9 & 10 \\
\midrule
$b^{(d)}$ 
& 40.326 & 43.594 & 44.961 & 46.473 & 46.731 
& 47.390 & 48.164 & 48.220 & 48.572 & 48.628 \\
ARL 
& \arlcell{50254.1}{556.5}
& \arlcell{50791.2}{527.2}
& \arlcell{49500.2}{628.8}
& \arlcell{50096.1}{603.2}
& \arlcell{49551.8}{505.1}
& \arlcell{49037.4}{550.1}
& \arlcell{49983.3}{589.9}
& \arlcell{50477.2}{463.1}
& \arlcell{49480.2}{412.0}
& \arlcell{50332.0}{574.9} \\
\midrule
\# selected ($d^\star=3$)
& 434 & 840 & \textbf{876} & 626 & 495 & 407 & 331 & 292 & 234 & 280 \\
\# selected ($d^\star=8$)
& 48 & 106 & 177 & 313 & 375 & 484 & 696 & \textbf{1101} & 840 & 664 \\
\bottomrule
\end{tabular}
}
\end{table}

Beyond identifying the true rank, the parallel strategy must also deliver competitive detection performance. We compare the proposed parallel procedure against the two fixed-rank approaches: MRS-C tuned to the true rank $d^\star$, and MRS-C with a misspecified rank that fixes $d = 1$. The fixed-rank methods are calibrated to achieve an ARL of 5000.

Detection performance is summarized in Table~\ref{tab:oracle_parallel_misspec}. Across both scenarios, the parallel strategy closely tracks the detector tuned to the true rank and substantially outperforms the misspecified fixed-rank alternative.
These results suggest that the parallel procedure handles rank uncertainty with only a modest increase in delay relative to the true-rank detector, while avoiding the larger loss caused by severe rank misspecification.

The counts of false alarms before the change-point $\tau$ in Table~\ref{tab:oracle_parallel_misspec} provide an empirical measure of the effective false-alarm rate.
Across both values of $d^\star$, the parallel procedure produces substantially fewer
premature alarms than the true-rank and misspecified alternatives, reflecting the
conservative nature of the Bonferroni calibration. In particular, although all methods are nominally calibrated to $\rm{ARL}\approx 5000$, the smaller number of early alarms for the parallel procedure indicates a larger effective ARL in finite samples. This conservatism partially offsets the slight increase in detection delay relative to the true-rank procedure and contributes to the overall robustness of the parallel approach.

Overall, these results demonstrate that the parallel MRS-C procedure provides a robust and practically effective alternative to a detector that knows the true rank, while avoiding the severe penalties associated with rank misspecification. The parallel procedure successfully controls false alarms and reliably detects the change without requiring prior knowledge of the true rank.

\begin{table}[t]
\centering
\caption{Comparison of true-rank, parallel, and misspecified MRS-C procedures under true ranks \(d^\star=3\) and \(d^\star=8\).
All methods are calibrated to achieve a global ARL of \(5000\) under the pre-change distribution.
Reported detection delays are computed over \(5000\) runs with \(T\ge \tau\), with standard errors shown in parentheses.
}
\label{tab:oracle_parallel_misspec}
\begin{tabular}{@{}c l c c@{}}
\toprule
True rank \(d^\star\) & Dimension selection & Detection delay & False alarms before \(\tau\) \\
\midrule
\multirow{3}{*}{\(3\)}
& True rank \((d=3)\)              & 101.80 (0.29) & 429 \\
& Parallel \((d\in\{1,\ldots,10\})\) & 107.21 (0.45) & 185 \\
& Misspecified \((d=1)\)           & 128.62 (0.62) & 399 \\
\midrule
\multirow{3}{*}{\(8\)}
& True rank \((d=8)\)              & 68.24 (0.13)  & 448 \\
& Parallel \((d\in\{1,\ldots,10\})\) & 69.99 (0.12)  & 196 \\
& Misspecified \((d=1)\)           & 102.15 (0.43) & 392 \\
\bottomrule
\end{tabular}
\end{table}

\subsection{Robotic Swarm Monitoring}\label{subsec:swarm_experiment}

In this section, we present robotic-swarm case studies to illustrate how the MRS-C statistic behaves on trajectory data with interpretable formation or motion transitions. We consider both a controlled synthetic swarm-trajectory dataset and a set of real-world flight logs from outdoor drone-flocking experiments. These two cases allow us to evaluate the method under complementary conditions: one in which the ground-truth change mechanisms are partially understood, and another in which changes are inferred solely from observed formation transitions.

\subsubsection{Synthetic milling data}\label{subsubsec:milling}

\begin{figure}[!tb]
    \centering
    \includegraphics[width=0.5\linewidth]{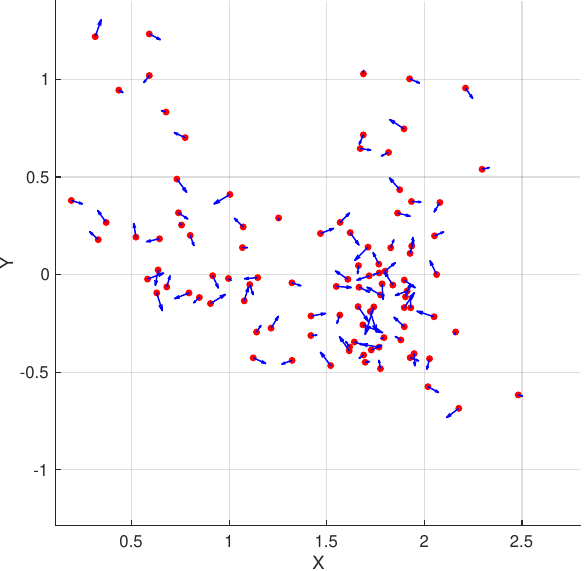}
    \caption{Snapshot of the synthetic milling swarm configuration at frame \(t=6001\). Red dots denote agent positions, and blue arrows indicate instantaneous velocity vectors.}
    \label{fig:snapshot}
\end{figure}

We consider a synthetic swarm-trajectory dataset comprising \(T=10001\) sequential frames of collective behavior, recorded every \(0.05\) s over the physical-time interval \(500.00\)--\(1000.00\) s. We index these frames as \(t=1,\ldots,T\). In each frame, there are \(n=100\) agents moving in a two-dimensional plane, and we record each agent's two-dimensional position and velocity. Let
\[
P_t=[p_{t1},\ldots,p_{tn}], \qquad
V_t=[v_{t1},\ldots,v_{tn}],
\]
where \(p_{ti},v_{ti}\in\Real^2\), \(i=1,\ldots,n\). Figure~\ref{fig:snapshot} shows a representative snapshot of the milling state at frame \(t=6001\). We construct an affinity graph in which each agent corresponds to a node, and we observe a time series of nodal features
\[
\widehat X_t =
\begin{bmatrix}
P_t\\
V_t
\end{bmatrix}
\in \Real^{4\times n}.
\]
Let \(\widetilde X_t\) be the position-centered and normalized nodal feature matrix obtained from \(\widehat X_t\). We vectorize $\widetilde X_t$ and use the resulting vector as the observation input $x_t$ for MRS-C.

\begin{figure}[!tb]
    \centering
    \includegraphics[width=0.8\linewidth]{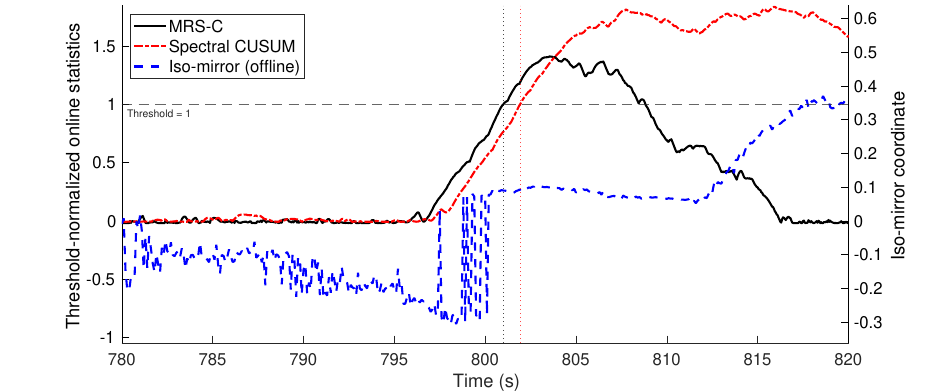}
    \caption{Normalized online monitoring statistics and the Iso-mirror \citep{chen2023discovering} coordinate over the transition interval of the swarm dataset. The online statistics, MRS-C and Spectral CUSUM, are normalized by their respective detection thresholds, so that the threshold level is one. The two online procedures signal within a similar time period as the retrospective Iso-mirror estimate.}
    \label{fig:results}
\end{figure}

We compare MRS-C with Spectral CUSUM~\citep{zhang2023spectral} as an online baseline. Spectral CUSUM monitors spectral summaries of time-varying affinity matrices through a CUSUM-type statistic. Here, the affinity matrices are constructed from the same normalized nodal features used by MRS-C. We also include an iso-mirror-based detector~\citep{chen2023discovering} as an offline retrospective reference, but not as an online competitor. The detailed definitions of the reference methods are provided in Appendix \ref{app:network_baselines}.

To make the online comparison fair, we calibrate MRS-C and Spectral CUSUM using the same stable milling segment from \(600\) to \(700\) s as nominal data, targeting the same empirical nominal ARL of \(500\) s, corresponding to \(10000\) frames. This is an empirical calibration based on the available nominal trajectory, not a model-based ARL guarantee, but it places the two online procedures on the same false-alarm scale.

Figure~\ref{fig:results} shows the normalized online monitoring statistics and the iso-mirror coordinate over the transition interval \(780\)--\(820\) s. MRS-C and Spectral CUSUM cross their thresholds at 801.00 s and 801.95 s, respectively, while the offline iso-mirror estimate exhibits a similar transition pattern. This agreement, under matched empirical nominal-ARL calibration for the two online methods, is consistent with MRS-C detecting a structural transition in the swarm dynamics.

\subsubsection{Real drone flocking flight logs}
\label{sec:real_drone_flocking}

We next evaluate MRS-C on real robotic-swarm trajectory data. For this purpose, we use the flight-log dataset associated with the 30-drone outdoor flocking experiments of \citet{vasarhelyi2018optimized}. The dataset consists of GPS-based GPX logs from four real swarm flights, denoted here by
\texttt{4mps\_diagonal},
\texttt{6mps\_circular},
\texttt{6mps\_obstacles}, and
\texttt{8mps\_circular}. Each flight contains trajectories from 30 autonomous drones moving collectively in a bounded outdoor environment. Our main goal is to assess whether the MRS-C statistic produces interpretable responses that align with independent physical summaries of the swarm's collective motion.

\paragraph*{Trajectory preprocessing.}
For each GPX log, we convert latitude, longitude, and altitude to local Cartesian coordinates using a local tangent-plane approximation. The trajectories from all drones in a given flight are then resampled to a common time grid with sampling interval \(\Delta t=0.2\) seconds over the time interval in which all 30 drones are observed. Let
\(
    p_{t,j}\in\mathbb{R}^3,\) \( j=1,\ldots,m,
\) denote the three-dimensional position of drone \(j\) at time index \(t\), where \(m=30\).

To focus on relative swarm geometry rather than global translation, we subtract the swarm centroid
\(
    \bar p_t=\sum_{j=1}^m p_{t,j}/m
\)
and scale by the root-mean-square swarm radius
\(
    r_t =
    \left\{\sum_{j=1}^m \|p_{t,j}-\bar p_t\|^2 / m
    \right\}^{1/2}.
\)
The normalized position of drone \(j\) is then
\(
    \widetilde p_{t,j}
    =
    (p_{t,j}-\bar p_t)/r_t.
\)
We also compute normalized finite-difference velocities
\(
    \widetilde v_{t,j}
    =
    (\widetilde p_{t,j}-\widetilde p_{t-1,j})/\Delta t.
\)
The feature vector \(x_t\) is obtained by vectorizing the normalized positions and velocities of all drones:
\[
    x_t =
    \operatorname{vec}
    \left(
    \widetilde p_{t,1},\ldots,\widetilde p_{t,m},
    \widetilde v_{t,1},\ldots,\widetilde v_{t,m}
    \right)
    \in \mathbb{R}^{180}.
\]
Feature standardization is performed using only a post-settling nominal reference window. Specifically, we use the interval \(250\)--\(450\) s as the nominal reference segment, estimate the coordinate-wise median and standard deviation from this segment, and then apply the resulting standardization to the full trajectory. This avoids using the initial takeoff and settling transient as the nominal reference behavior.

\paragraph*{MRS-C implementation.}
We apply MRS-C to the standardized feature sequence with window length \(w=25\) observations and subspace rank \(d=3\). Since \(\Delta t=0.2\) seconds, the subspace-estimation window corresponds to \(5\) seconds of data. We define the local growth score
\[
    G_t = S_t^{\mathrm{sub}} - S_{t-\ell}^{\mathrm{sub}},
\]
where \(\ell=25/\Delta t=125\) observations corresponds to a \(25\)-second window. The time at which \(G_t\) is maximized is reported as the period of strongest MRS-C growth.

\paragraph*{Physical order-parameter diagnostics.}
To interpret the MRS-C response, we compute several independent physical summaries of collective motion. These summaries are used as diagnostic quantities for interpreting the time periods highlighted by the monitoring statistic. Table~\ref{tab:physical_order_params} summarizes the quantities used in the analysis.

\begin{table}[t]
\centering
\small
\caption{{Physical order-parameter summaries used to interpret the real drone flight-log experiments. Here \(p_{t,j}\) and \(v_{t,j}\) denote the position and velocity of drone \(j\) at time \(t\), \(p^{xy}_{t,j}\) and \(v^{xy}_{t,j}\) denote their horizontal components, and \(m=30\).}}
\label{tab:physical_order_params}
\renewcommand{\arraystretch}{1.25}
\begin{tabular}{p{0.33\textwidth}p{0.67\textwidth}}
\toprule
{Summary }& {Definition}\\
\midrule
{Swarm radius
}&
{\(
\{m^{-1}\sum_j\|p_{t,j}-\bar p_t\|^2\}^{1/2}\)
}\\

{Mean NN distance
}&
{\(
m^{-1}\sum_j\min_{\ell\ne j}\|p_{t,j}-p_{t,\ell}\|\)
}\\

{Mean speed
}&
{\(
m^{-1}\sum_j\|v_{t,j}\|\)
}\\

{Aspect ratio
}&
{\(
(\lambda_1(\Gamma_t)/\lambda_2(\Gamma_t))^{1/2}\), where
\(\Gamma_t=\operatorname{Cov}\{p^{xy}_{t,j}\}_{j=1}^m\)
}\\

{Velocity polarization
}&
{\(
\left\|m^{-1}\sum_j v_{t,j}/\|v_{t,j}\|\right\|\)
}\\

{Milling order
}&
{\(
\left|
m^{-1}\sum_j
\frac{(p^{xy}_{t,j}-\bar p^{xy}_t)^\perp\cdot v^{xy}_{t,j}}{\|p^{xy}_{t,j}-\bar p^{xy}_t\|\,\|v^{xy}_{t,j}\|}
\right|\)
}\\
\bottomrule
\end{tabular}
\end{table}

We combine the standardized physical summaries into a vector $q_t$ and compute a retrospective local contrast
\[
    C_t =
    \left\|
    \frac{1}{\ell}\sum_{s=t+1}^{t+\ell} q_s
    -
    \frac{1}{\ell}\sum_{s=t-\ell+1}^{t} q_s
    \right\|_2,
\]
again using a 25-second window. 
The maximizer of $C_t$ after the nominal reference period is reported as the physical diagnostic change time. 
We note that this is an offline diagnostic used for interpretation only, and it is not a competing sequential stopping rule.

Table~\ref{tab:drone_flocking_results} summarizes the results across the four flights. 
The MRS-C growth peak is the time at which the local growth score $G_t$ is maximized. 
The physical diagnostic change time is the maximizer of the retrospective order-parameter contrast $C_t$.

\begin{table}[t]
\centering
\caption{Real 30-drone flight-log case study. The MRS-C growth peak is the time of maximum 25-second local growth of the MRS-C statistic. The physical diagnostic change time is computed from independent swarm-level order parameters.}
\label{tab:drone_flocking_results}
\begin{tabular}{@{}lrrr@{}}
\toprule
{Flight} 
& {MRS-C growth peak (s)}
& {Physical change (s)}
& {Lag (s)} \\
\midrule
{\texttt{4mps\_diagonal}}  
& {955.0}  
& {962.2}  
& {\(-7.2\)} \\
{\texttt{6mps\_circular}}  
& {1028.2} 
& {1036.6} 
& {\(-8.4\)} \\
{\texttt{6mps\_obstacles}} 
& {918.2}  
& {509.2}  
& {409.0} \\
{\texttt{8mps\_circular}}  
& {892.4}  
& {929.4}  
& {\(-37.0\)} \\
\bottomrule
\end{tabular}
\end{table}

\begin{figure}[!htb]
\centering
\includegraphics[width=\textwidth]{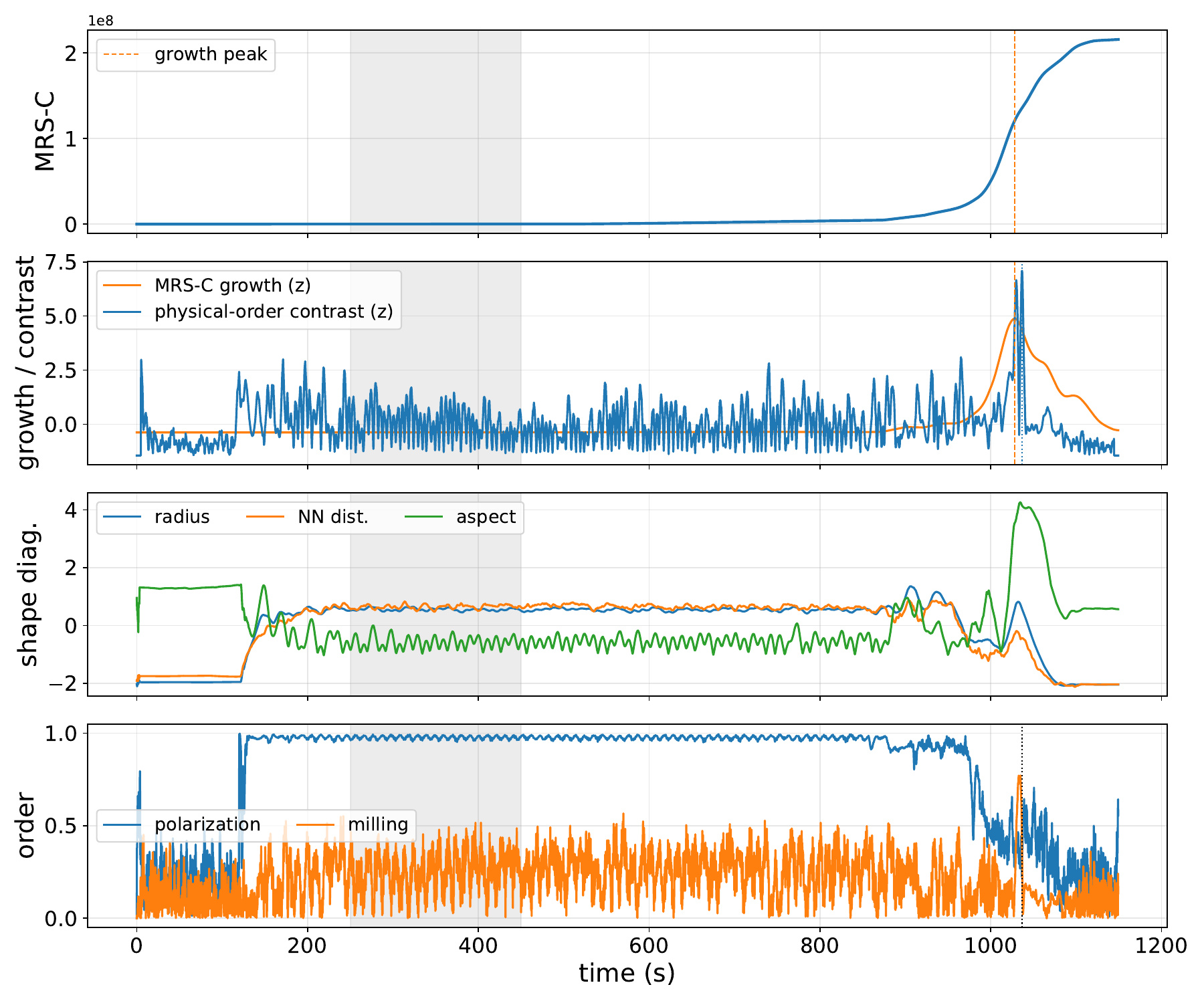}
\caption{ 
Real 30-drone flocking case study for the \texttt{6mps\_circular} flight. 
The top panel shows the MRS-C statistic. 
The second panel compares the 25-second local growth of the MRS-C statistic with a retrospective contrast computed from independent physical order parameters. 
The remaining panels show physical summaries of swarm geometry and collective motion. 
The strongest MRS-C growth occurs at 1028.2 seconds, close to the physical diagnostic change at 1036.6 seconds.}
\label{fig:drone_flocking_circular}
\end{figure}

Figure~\ref{fig:drone_flocking_circular} shows the diagnostic curves for the \texttt{6mps\_circular} flight, which provides the clearest representative example. 
The strongest local growth of the MRS-C statistic occurs at approximately 1028.2 seconds, close to the largest post-nominal physical-order change at approximately 1036.6 seconds. 
Around this period, the physical summaries indicate a substantial change in collective motion. The velocity polarization drops sharply, the aspect ratio changes substantially, and the nearest-neighbor and radius summaries also shift. 
This temporal agreement suggests that the subspace statistic is responding to a collective reconfiguration of the swarm rather than to isolated trajectory noise.

The \texttt{6mps\_obstacles} flight exhibits a more complex pattern. The physical diagnostic identifies an early disturbance around \(509.2\) seconds, whereas the largest MRS-C growth occurs later, around \(918.2\) seconds. This pattern is consistent with a prolonged or multi-stage disturbance rather than a single abrupt reconfiguration. Such behavior is plausible in the obstacle setting, where interactions with the environment may induce several phases of collective adaptation.

Overall, this real-flight study illustrates that MRS-C can be applied end-to-end to high-dimensional robotic-swarm trajectory data and can produce monitoring signals that are interpretable in terms of independent physical summaries of collective motion.

\section{CONCLUSION}\label{sec:conclusion}

    In this paper, we present the \emph{Multi-rank Subspace-CUSUM} (MRS-C) procedure for the sequential detection of changes in covariance structure in high-dimensional streams. Our method generalizes prior subspace monitoring approaches to handle multiple concurrent spikes, thereby enhancing detection capabilities in scenarios with multiple simultaneous low-rank signals. We established a solid theoretical foundation to guide the selection of key tuning parameters that control the false-alarm rate and minimize detection delay. The analysis also yields an oracle-relative efficiency characterization, with a constant that captures the effect of heterogeneous spike strengths.

    We also addressed the practical challenge of unknown subspace dimensions through a parallel monitoring scheme. Synthetic experiments support the theoretical scaling and illustrate the operating regimes of MRS-C relative to the baseline methods. Robotic-swarm case studies show that the proposed statistic can identify interpretable formation changes in trajectory data. Future work may extend this framework to non-Gaussian settings or time-varying covariances and explore adaptive strategies for dynamic subspace dimension selection.

\section*{FUNDING}
This work is supported by the Office of Naval Research [N000142412278]. Ira B. Schwartz and Jason Hindes were supported by the Office of Naval Research [N0001425GI101158] and NRL Base funding [N0001424WX00013].

\section*{DISCLOSURE OF INTEREST}
No potential competing interest was reported by the authors.

\appendix
\setcounter{table}{0}
\setcounter{figure}{0}
\setcounter{equation}{0}

\renewcommand{\thetable}{A.\arabic{table}}
\renewcommand{\thefigure}{A.\arabic{figure}}
\renewcommand{\theequation}{A.\arabic{equation}}
\renewcommand{\thealgorithm}{A.\arabic{algorithm}}

\section{PROOFS}\label{app:proofs}

Proofs of results presented in Section~\ref{sec:analysis} are provided below.

\begin{proof}[Proof of Lemma \ref{lem:post_change_increment}]
For notational simplicity, we here denote by $\hat{u}_i$, instead of $\hat u_{i, t+w}$, the $i$-th leading unit-norm eigenvector of the sample covariance $\hat \Sigma_{t+w}$, which is the $i$th column of $\hat U_{t+w}$.
Let $\{u_{d+1},\dots,u_k\}$ be an arbitrary orthonormal basis of $\mathrm{col}(U)^\perp$.
Under the spiked covariance model with distinct signal strengths $\{\lambda_i\}$, we have the following asymptotic normality of $\hat u_i$, $i = 1,\dots, d$ \citep{anderson1963asymptotic, paul2007asymptotics}:
\begin{align}\label{eq:eig_asymptotic}
    \sqrt{w}(\hat u_i - u_i) \stackrel{d} \to N(0, \Xi_i),
\end{align}
where the $k \times k$ covariance matrix $\Xi_i$ admits the spectral decomposition as
\[
    \Xi_i
    \;=\;
    \sum_{\substack{j=1\\j\neq i}}^k
    \frac{\mu_i\mu_j}{(\mu_i-\mu_j)^2}\;u_j\,u_j^T,
    \quad
    \mu_j =
    \begin{cases}
    \sigma^2 + \lambda_j, & j = 1,\dots,d,\\
    \sigma^2, & j = d+1,\dots,k.
    \end{cases}
\]
This $\Xi_i$ can be rewritten as
\begin{align}\label{eq:Xi_i}
    \Xi_i = \frac{1 + \rho_i}{\rho_i^2}(I_k - UU^T) + \sum_{\substack{j=1\\j \neq i}}^d \frac{(1+\rho_i)(1+\rho_j)}{(\rho_i - \rho_j)^2} u_j u_j^T.
\end{align}
By the eigenvector CLT \eqref{eq:eig_asymptotic}, we may write
\[
\hat u_i
= u_i + \frac{1}{\sqrt w} \xi_i + o_p\!\left(\frac{1}{\sqrt w}\right),
\]
where $\xi_i \sim N(0,\Xi_i)$ is orthogonal with $u_i$.
Consider the normalization map $g(v)=v/\|v\|$.
A second-order Taylor expansion of $g$ around a unit vector $u$ yields
\[
g(u+h)
= u + (I-uu^T)h
- \frac{1}{2}\big\|(I-uu^T)h\big\|^2 u
+ o(\|h\|^2).
\]
Taking $h = w^{-1/2} \xi_i$ and using $u_i^T \xi_i = 0$, we obtain
\[
\hat u_i
= u_i + \frac{1}{\sqrt w} \xi_i
- \frac{1}{2w}\|\xi_i\|^2 u_i
+ o_p\!\left(\frac{1}{w}\right).
\]
Consequently,
\[
\hat u_i \hat u_i^T
= u_i u_i^T
+ \frac{1}{\sqrt w}\big(u_i \xi_i^T + \xi_i u_i^T\big)
+ \frac{1}{w}\!\left(\xi_i \xi_i^T - \|\xi_i\|^2 u_i u_i^T\right)
+ o_p\!\left(\frac{1}{w}\right),
\]
and hence
\begin{align}\label{eq:expec_ui_cross}
\mathbb{E}_0(\hat u_i \hat u_i^T)
= u_i u_i^T
+ \frac{1}{w}\!\left(\Xi_i - \operatorname{tr}(\Xi_i)\,u_i u_i^T\right)
+ o\!\left(\frac{1}{w}\right).
\end{align}
From \eqref{eq:Xi_i}, we have
\begin{align}\label{eq:Xi_trace}
\tr(\Xi_i) = (k-d) \frac{1+\rho_i}{\rho_i^2} + \sum_{\substack{j=1\\j \neq i}}^d \frac{(1+\rho_i) (1+\rho_j)}{(\rho_i - \rho_j)^2}.
\end{align}
Recall that $\hat u_i$ is independent of $x_t$.
Using \eqref{eq:expec_ui_cross} and \eqref{eq:Xi_trace}, we get
\begin{align*}
    \E_0(\hat u_i^T x_t)^2 &= \E_0\big[ \hat u_i^T \E_0[x_t x_t^T] \hat u_i \big]\\
    &=\tr \big[ (\sigma^2 I_k + U \Lambda U^T) \E_0(\hat u_i \hat u_i^T)\big]\\
    &=\tr \Big[ (\sigma^2 I_k + U \Lambda U^T) \Big\{\frac{1}{w}\Xi_i + \Big(1 - \frac{\tr(\Xi_i)}{w} \Big) u_i u_i^T \Big\}\Big] + o\Big(\frac{1}{w}\Big)\\
    &=\tr \Big[ (\sigma^2 I_k + U \Lambda U^T) \Big\{\frac{1+\rho_i}{w\rho_i^2}(I_k - U U^T) + \frac{1}{w} \sum_{\substack{j=1\\j \neq i}}^d \frac{(1+\rho_i)(1+\rho_j)}{(\rho_i - \rho_j)^2} u_j u_j^T
    \\ & \qquad + \Big(1 - \frac{1+\rho_i}{w \rho_i^2} (k-d) - \frac{1}{w} \sum_{\substack{j=1\\j \neq i}}^d \frac{(1+\rho_i)(1+\rho_j)}{(\rho_i - \rho_j)^2} \Big) u_i u_i^T \Big\}\Big] + o\Big(\frac{1}{w}\Big)\\
    &=\sigma^2 (1 + \rho_i) \Big\{ 1 - \frac{k-d}{w \rho_i} - \frac{1}{w} \sum_{\substack{j=1\\j \neq i}}^d \frac{1+\rho_j}{\rho_i - \rho_j} \Big\} + o\Big(\frac{1}{w}\Big).
\end{align*}
Therefore, summing over $i$ from $1$ to $d$, the double sum vanishes by antisymmetry, and we obtain the expected post-change increment by
\begin{align*}
\begin{split}
    \E_0 Z_t &= \sum_{i=1}^d \E_0 (\hat u_i^T x_t)^2\\
    &= \sigma^2 \sum_{i=1}^d (1+\rho_i) \Big(1- \frac{k-d}{w\rho_i}\Big) - \frac{\sigma^2}{w} \sum_{i=1}^d \sum_{\substack{j=1\\j \neq i}}^d \frac{(1+\rho_i)(1+\rho_j)}{\rho_i - \rho_j} + o\Big(\frac{1}{w}\Big)\\
    &= \sigma^2 \sum_{i=1}^d (1+\rho_i) \Big(1- \frac{k-d}{w\rho_i}\Big) + o\Big(\frac{1}{w}\Big).
\end{split}
\end{align*}
\end{proof}

We now prove Theorem~\ref{thm:arl-edd}. We begin with the following lemma, which establishes a concentration bound for the post-change increments. Let
\[
    Y_t = Z_t - \Delta, \quad
    \mu_w = \E_0 Y_t, \quad
    \hat P_{t+w} = \hat U_{t+w} \hat U_{t+w}^T.
\]

\begin{lem}
\label{lem:wdep-concentration}
Under $\P_0$, $\{Y_t\}_{t\ge1}$ is stationary and $w$-dependent.
Moreover, for any interval $I\subset\mathbb N$ of length $n$, define
\[
    V_I=\sum_{t\in I}(Y_t-\mu_w).
\]
There exists a constant $c_1>0$, independent of $w, n, I, \gamma$, such that
\[
    \mathbb P_0(|V_I|>x)
    \le
    2\exp\left[
        -c_1
        \min\left\{
            \frac{x^2}{n(w+1)},
            \frac{x}{w+1}
        \right\}
    \right].
\]
\end{lem}

\begin{proof}
For each $t$, $Y_t$ is a measurable function of the data
block
\(
    (x_t,x_{t+1},\ldots,x_{t+w}).
\)
If $|s-t|>w$, the corresponding raw-data blocks are disjoint. Since the
observations are i.i.d. under $\mathbb P_0$, the two increments are
independent. Hence $\{Y_t\}$ is $w$-dependent.
Stationarity follows from the stationarity of the post-change observations.

Let
\[
\hat P_{t+w} \coloneqq \hat U_{t+w} \hat U_{t+w}^T, \quad
\Sigma_0=\sigma^2I_k+U\Lambda U^{T}, \quad B_t=\Sigma_0^{1/2}\hat P_{t+w}\Sigma_0^{1/2}.
\] 
Since $\hat P_{t+w}$ is an orthogonal projection of rank $d$,
\[
    \left\|
        B_t
    \right\|_{\rm op}
    \le
    \|\Sigma_0\|_{\rm op}
    =
    \sigma^2+\lambda_1 \eqqcolon M.
\]
Thus, conditional on $\hat P_{t+w}$,
\[
    Z_t\overset{d}{=}\sum_{j=1}^{r_t}a_{j,t}\xi_j^2,
    \qquad
    r_t\le d,\quad 0\le a_{j,t}\le M,
\]
where $\xi_j\stackrel{\rm i.i.d.}{\sim}N(0,1)$. Hence, for
$|\eta|\le (4M)^{-1}$,
\[
\begin{aligned}
    \mathbb E_0\left[
        e^{\eta(Z_t-\operatorname{tr}B_t)}
        \mid \widehat P_{t+w}
    \right]
    &=
    \prod_{j=1}^{r_t}
    e^{-\eta a_{j,t}}
    (1-2\eta a_{j,t})^{-1/2}  \\
    &\le
    \exp\left\{
        C\eta^2\sum_{j=1}^{r_t}a_{j,t}^2
    \right\} 
    \le
    \exp(CdM^2\eta^2).
\end{aligned}
\]
Moreover,
\(
    0\le \operatorname{tr}B_t\le dM,
\)
so by Hoeffding's lemma,
\[
    \mathbb E_0
    e^{\eta(\operatorname{tr}B_t-\mathbb E_0Z_t)}
    \le
    \exp(Cd^2M^2\eta^2).
\]
Therefore, using the tower property,
for $|\eta|\le (4M)^{-1}$,
\[
\begin{aligned}
    \mathbb E_0 e^{\eta(Y_t-\mu_w)} = \mathbb E_0 e^{\eta(Z_t-\mathbb E_0Z_t)}
    &=
    \mathbb E_0\left[
        e^{\eta(\operatorname{tr}B_t-\mathbb E_0Z_t)}
        \mathbb E_0
        \left\{
            e^{\eta(Z_t-\operatorname{tr}B_t)}
            \mid \widehat P_{t+w}
        \right\}
    \right] \\
    &\le
    \exp(C_0\eta^2),
\end{aligned}
\]
where $C_0$ is independent of $t$ or $w$.

Now partition the interval $I$ into $w+1$ residue classes modulo $w+1$.
Within each residue class, the variables are independent. By
H\"older's inequality, for $|\eta|\le(4M)^{-1}/(w+1)$,
\[
\begin{aligned}
    \mathbb E_0 e^{\eta V_I}
    &=
    \mathbb E_0
    \exp\left\{
        \eta\sum_{r=0}^{w}
        \sum_{\substack{t\in I\\ t\equiv r\,({\rm mod}\,w+1)}}
        (Y_t-\mu_w)
    \right\}  \\
    &\le
    \prod_{r=0}^{w}
    \left[
        \mathbb E_0
        \exp\left\{
            (w+1)\eta
            \sum_{\substack{t\in I\\ t\equiv r\,({\rm mod}\,w+1)}}
            (Y_t-\mu_w)
        \right\}
    \right]^{1/(w+1)}  \\
    &\le
    \exp\{C_0 n(w+1)\eta^2\}.
\end{aligned}
\]
Chernoff's bound, applied to $V_I$ and to $-V_I$, yields
\[
    \mathbb P_0(|V_I|>x)
    \le
    2\exp\left[
        -c_1
        \min\left\{
            \frac{x^2}{n(w+1)},
            \frac{x}{w+1}
        \right\}
    \right].
\]
\end{proof}

\begin{proof}[Proof of Theorem~\ref{thm:arl-edd}]

We first handle the pre-change calibration. Under $\mathbb P_\infty$,
\(
    x_t\stackrel{\rm i.i.d.}{\sim}N(0,\sigma^2I_k).
\)
Let
\(
    \mathcal G_{t+1}=\sigma(x_{t+1},x_{t+2},\ldots).
\)
Conditional on $\mathcal G_{t+1}$, the matrix $\widehat P_{t+w}$ is fixed
and is an orthogonal projection of rank $d$, while $x_t$ is independent of
$\mathcal G_{t+1}$. Hence, by the rotational invariance of the Gaussian
distribution,
\[
    Z_t\mid \mathcal G_{t+1}
    =
    x_t^T\widehat P_{t+w}x_t\mid\mathcal G_{t+1}
    \sim
    \sigma^2\chi_d^2 .
\]
This conditional distribution does not depend on the realized value of
$\widehat P_{t+w}$. Thus, for any bounded measurable function $h$,
\[
    \mathbb E_\infty\{h(Z_t)\mid\mathcal G_{t+1}\}
    =
    \mathbb E\{h(\sigma^2\chi_d^2)\},
\]
which is deterministic. Therefore $Z_t$ is independent of
$\mathcal G_{t+1}$. A backward-conditioning argument then shows that, under $\P_\infty$, $Z_t$ are independently distributed $\sigma^2\chi_d^2$.
Consequently, given $\delta \in (0, (2\sigma^2)^{-1})$,
\(
    \mathbb E_\infty e^{\delta(Z_t-\Delta)}
    =
    e^{-\delta\Delta}(1-2\sigma^2\delta)^{-d/2}.
\)
Choose
\[
    \Delta
    = \Delta(\delta) = 
    -\frac{d}{2\delta}
    \log(1-2\sigma^2\delta),
\]
so that
\(
    \mathbb E_\infty e^{\delta(Z_t-\Delta)}=1.
\)
Since the pre-change increments are i.i.d.~with negative drift, the standard calibration for the reflected random walk gives
\(
    b_\gamma
    = b_\gamma(\delta)=
\log\gamma/\delta+O(1),
\)
for a threshold calibrated to ARL level $\gamma$.

We now analyze the post-change delay. By
Lemma~\ref{lem:wdep-concentration}, under $\mathbb P_0$, the sequence
$\{Y_t\}$ is stationary and $w$-dependent.
By Lemma~\ref{lem:post_change_increment},
\[
    \mathbb E_0[Z_t]
    =
    \sigma^2A(w)+o(1/w),
\]
so
\(
    \mu_w
    =
    \sigma^2A(w)-\Delta+o(1/w).
\)
Let
\(
    \nu_b=\inf\{t\ge1:S_t^{\rm sub}\ge b\}.
\)
Then
\(
    T^{\rm sub}=\nu_{b_\gamma}+w.
\)
The recursion admits the pathwise representation
\(
    S_n^{\rm sub}
    =
    \max_{0\le j<n}
    \sum_{t=j+1}^{n}Y_t.
\)

We next show that
\(
    \mathbb E_0[\nu_{b_\gamma}]
    =
    b_\gamma/\mu_w\{1+o(1)\}.
\)
First, we prove the lower bound. Fix $\varepsilon\in(0,1)$ and define
\(
    n_-=
    \left\lfloor
        (1-\varepsilon)b_\gamma/\mu_w
    \right\rfloor.
\)
If $\nu_{b_\gamma}\le n_-$, then for some $0\le j<n\le n_-$,
\(
    \sum_{t=j+1}^{n}Y_t\ge b_\gamma.
\)
For such an interval,
\[
    (n-j)\mu_w\le n_-\mu_w\le(1-\varepsilon)b_\gamma,
\]
so
\(
    \sum_{t=j+1}^{n}(Y_t-\mu_w)\ge\varepsilon b_\gamma.
\)
There are at most $n_-^2=O(b_\gamma^2)$ candidate intervals. By
Lemma~\ref{lem:wdep-concentration},
\[
    \mathbb P_0(\nu_{b_\gamma}\le n_-)
    \le
    Cb_\gamma^2
    \exp\left(-c\frac{b_\gamma}{w}\right)
    =
    o(1),
\]
because $b_\gamma\asymp\log\gamma$ and
\(
    (w\log b_\gamma)/b_\gamma\to0.
\)
Thus
\(
    \mathbb P_0(\nu_{b_\gamma}>n_-)\to1,
\)
and consequently
\[
    \mathbb E_0[\nu_{b_\gamma}]
    \ge
    \E_0[\nu_{b_\gamma} \1\{\nu_{b_\gamma} \ge n_-\}]
    \ge
    n_-\mathbb P_0(\nu_{b_\gamma}>n_-)
    =
    (1-\varepsilon)\frac{b_\gamma}{\mu_w}\{1+o(1)\}.
\]
Letting $\varepsilon\downarrow0$ yields
\begin{align}\label{eq:edd_lower_bound}
    \liminf_{\gamma\to\infty}
    \frac{\mathbb E_0[\nu_{b_\gamma}]}{b_\gamma/\mu_w}
    \ge1.
\end{align}
We now prove the upper bound. Fix $\varepsilon>0$ and define
\(
    n_+
    =
    \left\lceil
        (1+\varepsilon)b_\gamma/\mu_w
    \right\rceil.
\)
By Lemma~\ref{lem:wdep-concentration}, applied to the interval
$\{1,\ldots,n_+\}$,
\[
    q_\gamma
    :=
    \mathbb P_0\left(
        \sum_{t=1}^{n_+}Y_t<b_\gamma
    \right)
    \le
    C\exp\left(-c\frac{b_\gamma}{w}\right)
    =
    o(1).
\]
Set
\(
    L=n_+ + w.
\)
For $r=1,2,\ldots$, define
\(
    a_r=(r-1)L+1,
\)
and the success events
\[
    E_r
    =
    \left\{
        \sum_{t=a_r}^{a_r+n_+-1}Y_t\ge b_\gamma
    \right\}.
\]
The event $E_r$ depends only on the observations
\(
    \{x_{a_r},x_{a_r+1},\ldots,x_{a_r+n_++w-1}\}.
\)
Since consecutive starting times are separated by $L=n_+ + w$, these
raw-data blocks are disjoint for different $r$'s. Hence
$E_1,E_2,\ldots$ are independent. By stationarity,
\(
    \mathbb P_0(E_r)=1-q_\gamma,\) \(r=1,2,\ldots .
\)
Let
\(
    R=\inf\{r\ge1:E_r\text{ occurs}\}.
\)
Then $R$ is geometric with success probability $1-q_\gamma$, and
\(
    \mathbb E_0[R]=1/({1-q_\gamma}).
\)

If $E_r$ occurs, then the pathwise representation of $S_t$ gives
\(
    S_{a_r+n_+-1}
    \ge
    \sum_{t=a_r}^{a_r+n_+-1}Y_t
    \ge b_\gamma.
\)
Therefore
\(
    \nu_{b_\gamma}
    \le
    a_R+n_+-1
    =
    (R-1)L+n_+ .
\)
Taking expectations,
\[
    \mathbb E_0[\nu_{b_\gamma}]
    \le
    L\{\mathbb E_0[R]-1\}+n_+
    =
    L\frac{q_\gamma}{1-q_\gamma}+n_+ .
\]
Since $q_\gamma\to0$, $L=O(b_\gamma)$, and
\(
    Lq_\gamma
    =
    O(b_\gamma)
    \exp\left(-c{b_\gamma}/{w}\right)
    =
    o(b_\gamma),
\)
we obtain
\[
    \mathbb E_0[\nu_{b_\gamma}]
    \le
    n_+ + o(b_\gamma)
    =
    (1+\varepsilon)\frac{b_\gamma}{\mu_w}\{1+o(1)\}.
\]
Letting $\varepsilon\downarrow0$ gives
\begin{align}\label{eq:edd_upper_bound}
    \limsup_{\gamma\to\infty}
    \frac{\mathbb E_0[\nu_{b_\gamma}]}{b_\gamma/\mu_w}
    \le1.
\end{align}
Combining the lower bound \eqref{eq:edd_lower_bound} and the upper bound \eqref{eq:edd_upper_bound}, we have the desired results.
Finally,
\begin{align}
    \nonumber \mathbb E_0[T_{\rm sub}] &=
    \mathbb E_0[\nu_{b_\gamma}]+w =
    \nonumber \frac{b_\gamma}{\mu_w}\{1+o(1)\}+w =
    \nonumber \frac{\log\gamma}{\delta\mu_w}
    \{1+o(1)\}+w \\&=
    \label{eq:edd_delta} \frac{\log \gamma}{\delta (\E_0 Z_t - \Delta(\delta))}\{1 + o(1)\} + w.
\end{align}
We now choose $\delta$ as the unique maximizer of the denominator:
\[
    \delta_\infty
    =
    \left(1-{d}/{A(w)}\right)/{(2\sigma^2)}.
\]
The corresponding drift is
\[
    \Delta
    =
    \Delta(\delta_\infty)
    =
    \frac{d\sigma^2}{1-d/A(w)}
    \log\left(\frac{A(w)}{d}\right).
\]
Substituting these into \eqref{eq:edd_delta} gives the final result.
\end{proof}

\begin{proof}[Proof of Theorem \ref{thm:optimality}]
    Recall from \eqref{eq:KL_zero} that the Kullback-Leibler divergence is given by
    \begin{align*}
        \Ic_0 \coloneqq \E_{0}\log\left(\frac{f_0}{f_\infty}\right) = \frac{1}{2}\sum_{i=1}^d [\rho_i - \log(1+\rho_i)].
    \end{align*}
    Therefore, by the results in \cite{tartakovsky2014sequential}, we have
    \begin{align}
        \E_0[T_C] = \frac{\log \gamma}{\Ic_0} + O(1) = \frac{2 \log \gamma}{\sum_{i=1}^d [\rho_i - \log(1+\rho_i)]} + O(1).
    \end{align}
    Using the result from  Theorem \ref{thm:arl-edd} and the optimal window size $w^* \asymp \sqrt{\log \gamma}$ from Corollary \ref{thm:opt_window}, the EDD of MRS-C is
    \begin{align*}
        {\E_0 }[{T^{\mathrm{sub}}}] = \frac{2 \log \gamma}{A(w^*) - d (1 + \log(A(w^*)/d))} + w^* + O(1),
    \end{align*}
    and the denominator term approaches its limit
    \begin{align*}
        {D^\infty \coloneqq \lim_{\gamma \to \infty} \left[ A(w^*) - d \left(1 + \log\Big(\frac{A(w^*)}{d}\Big)\right) \right]= \sum_{i=1}^d }[{\rho_i - \log (1+\bar \rho)}],
    \end{align*}
    with an error of order $O(1/\sqrt{\log \gamma})$. Therefore, the EDD for MRS-C is expanded as
    \begin{align}
        \E_0[T^{\mathrm{sub}}] = \frac{2 \log \gamma}{\sum_{i=1}^d [\rho_i - \log (1+\bar \rho)]} + O\big(\sqrt{\log \gamma}\big).
    \end{align}
    Using the geometric series expansion $(1+x)^{-1} = 1 - x + O(x^2)$, we get
    \begin{align*}
        \frac{\E_0[T^{\mathrm{sub}}]}{\E_0[T_C]} &= \left(\frac{2\Ic_0}{D^\infty} + O\Big(\frac{1}{\sqrt{\log \gamma}}\Big)\right) \left(1 - O\Big(\frac{1}{\log \gamma}\Big)\right)\\
        &= \Kc + O \Big( \frac{1}{\sqrt{\log \gamma}} \Big).
    \end{align*}
    Since $x \mapsto \log (1+x)$ is strictly concave, Jensen's inequality states that
    \begin{align*}
        \frac{1}{d} \sum_{i=1}^d \log(1+\rho_i) \le \log(1 + \bar\rho).
    \end{align*}
    Therefore, $\Kc \ge 1$, with equality holding if and only if all $\rho_i$ are equal.
\end{proof}

\section{ADDITIONAL EXPERIMENTS}\label{app:experiment_details}
\subsection{Comprehensive simulation results}\label{app:simul_comprehensive}

We provide comprehensive simulation results for spiked covariance scenarios mentioned in Section \ref{subsec:simulations}. Tables \ref{tab:sim_1_dense} and \ref{tab:sim_1_sparse} are results for a rank-1 spike model with a dense and sparse signal direction, respectively. Table \ref{tab:sim_d_nonuniform} shows the results for a multi-rank spike model with non-uniform signal strengths.

\begin{table}[!h]
\centering
\caption{Rank-1 dense spike: $\text{ARL} = 5000$, $\lambda = 1$, $u = (1,1,\dots,1)^T / \sqrt{k}$.}
\label{tab:sim_1_dense}
\small
\setlength{\tabcolsep}{4pt}
\renewcommand{\arraystretch}{1.05}
\resizebox{\textwidth}{!}{%
\begin{tabular}{cc
    *{9}{>{\centering\arraybackslash}p{0.085\textwidth}}}
\toprule
\multirow{2}{*}{$k$} & \multirow{2}{*}{$d$}
  & \multicolumn{3}{c}{$\sigma^2=2$}
  & \multicolumn{3}{c}{$\sigma^2=1$}
  & \multicolumn{3}{c}{$\sigma^2=0.5$}
\\
\cmidrule(lr){3-5} \cmidrule(lr){6-8} \cmidrule(lr){9-11}
&
  & CUSUM & MRS-C & LESC
  & CUSUM & MRS-C & LESC
  & CUSUM & MRS-C & LESC
\\
\midrule
5 & 1
&  90.1 \scriptsize($ 0.84$) & 211.4 \scriptsize($ 3.36$) & 550.3 \scriptsize($ 7.53$)
&  35.5 \scriptsize($ 0.27$) &  97.7 \scriptsize($ 1.98$) & 136.2 \scriptsize($ 1.45$)
&  14.7 \scriptsize($ 0.13$) &  65.6 \scriptsize($ 0.27$) &  57.4 \scriptsize($ 0.28$)
\\
10 & 1
&  90.6 \scriptsize($ 1.23$) & 303.2 \scriptsize($ 7.42$) & 909.7 \scriptsize($ 30.04$)
&  35.1 \scriptsize($ 0.30$) & 119.1 \scriptsize($ 1.92$) & 167.1 \scriptsize($ 1.16$)
&  14.7 \scriptsize($ 0.21$) &  74.3 \scriptsize($ 0.71$) &  66.4 \scriptsize($ 0.41$)
\\
20 & 1
&  90.5 \scriptsize($ 0.90$) & 644.1 \scriptsize($ 28.73$) & 2194.4 \scriptsize($ 66.08$)
&  35.5 \scriptsize($ 0.12$) & 136.4 \scriptsize($ 1.59$) & 203.5 \scriptsize($ 3.34$)
&  14.7 \scriptsize($ 0.14$) &  80.1 \scriptsize($ 1.03$) &  72.9 \scriptsize($ 0.39$)
\\
\bottomrule
\end{tabular}%
}
\end{table}

\begin{table}[!h]
\centering
\caption{Rank-1 sparse spike: $\text{ARL} = 5000$, $\lambda = 1$, $u = e_1$.}
\label{tab:sim_1_sparse}
\small
\setlength{\tabcolsep}{4pt}
\renewcommand{\arraystretch}{1.05}
\resizebox{\textwidth}{!}{%
\begin{tabular}{cc
    *{9}{>{\centering\arraybackslash}p{0.085\textwidth}}}
\toprule
\multirow{2}{*}{$k$} & \multirow{2}{*}{$d$}
  & \multicolumn{3}{c}{$\sigma^2=2$}
  & \multicolumn{3}{c}{$\sigma^2=1$}
  & \multicolumn{3}{c}{$\sigma^2=0.5$}
\\
\cmidrule(lr){3-5} \cmidrule(lr){6-8} \cmidrule(lr){9-11}
&
  & CUSUM & MRS-C & LESC
  & CUSUM & MRS-C & LESC
  & CUSUM & MRS-C & LESC
\\
\midrule
5 & 1
&  90.9 \scriptsize($ 0.91$) & 230.6 \scriptsize($ 5.44$) & 561.0 \scriptsize($ 5.91$)
&  35.2 \scriptsize($ 0.39$) & 102.4 \scriptsize($ 2.11$) & 141.2 \scriptsize($ 0.43$)
&  14.8 \scriptsize($ 0.18$) &  69.5 \scriptsize($ 0.52$) &  59.9 \scriptsize($ 0.28$)
\\
10 & 1
&  90.3 \scriptsize($ 1.12$) & 411.5 \scriptsize($ 8.35$) & 911.4 \scriptsize($ 15.25$)
&  35.6 \scriptsize($ 0.35$) & 112.8 \scriptsize($ 2.07$) & 161.2 \scriptsize($ 0.74$)
&  14.9 \scriptsize($ 0.24$) &  70.3 \scriptsize($ 0.94$) &  62.1 \scriptsize($ 0.30$)
\\
20 & 1
&  90.4 \scriptsize($ 1.05$) & 788.7 \scriptsize($ 34.20$) & 2255.1 \scriptsize($ 52.51$)
&  35.4 \scriptsize($ 0.21$) & 145.8 \scriptsize($ 2.40$) & 198.4 \scriptsize($ 1.98$)
&  14.8 \scriptsize($ 0.19$) &  73.0 \scriptsize($ 1.38$) &  68.7 \scriptsize($ 0.26$)
\\
\bottomrule
\end{tabular}%
}
\end{table}

\begin{table}[!h]
\centering
\caption{Rank-$d$ non-uniform spikes: $\text{ARL} = 5000$, $\Lambda = \operatorname{diag}(\operatorname{linspace}(2,1,d))$.}
\label{tab:sim_d_nonuniform}
\small
\setlength{\tabcolsep}{4pt}
\renewcommand{\arraystretch}{1.05}
\resizebox{\textwidth}{!}{%
\begin{tabular}{cc
    *{9}{>{\centering\arraybackslash}p{0.085\textwidth}}}
\toprule
\multirow{2}{*}{$k$} & \multirow{2}{*}{$d$}
  & \multicolumn{3}{c}{$\sigma^2=2$}
  & \multicolumn{3}{c}{$\sigma^2=1$}
  & \multicolumn{3}{c}{$\sigma^2=0.5$}
\\
\cmidrule(lr){3-5} \cmidrule(lr){6-8} \cmidrule(lr){9-11}
&
  & CUSUM & MRS-C & LESC
  & CUSUM & MRS-C & LESC
  & CUSUM & MRS-C & LESC
\\
\midrule
5  & 2
&  28.6 \scriptsize($ 0.15$) &  90.5 \scriptsize($ 0.57$) &  324.7 \scriptsize($ 4.65$)
&  11.6 \scriptsize($ 0.11$) &  63.9 \scriptsize($ 0.17$) &   67.2 \scriptsize($ 0.46$)
&   5.3 \; \scriptsize($ 0.02$) &  56.0 \scriptsize($ 0.05$) &   38.9 \scriptsize($ 0.18$)
\\
5  & 3
&  20.8 \scriptsize($ 0.11$) &  76.9 \scriptsize($ 0.32$) &  256.6 \scriptsize($ 2.51$)
&   8.4 \; \scriptsize($ 0.04$) &  61.0 \scriptsize($ 0.10$) &   63.8 \scriptsize($ 1.25$)
&   3.9 \; \scriptsize($ 0.00$) &  54.2 \scriptsize($ 0.05$) &   35.5 \scriptsize($ 0.25$)
\\
10 & 2
&  28.5 \scriptsize($ 0.23$) & 109.4 \scriptsize($ 1.10$) &  581.1 \scriptsize($ 2.60$)
&  11.6 \scriptsize($ 0.03$) &  65.5 \scriptsize($ 0.27$) &   76.5 \scriptsize($ 1.95$)
&   5.3 \; \scriptsize($ 0.03$) &  56.7 \scriptsize($ 0.08$) &   43.2 \scriptsize($ 0.20$)
\\
10 & 3
&  20.8 \scriptsize($ 0.09$) &  90.9 \scriptsize($ 0.53$) &  403.0 \scriptsize($ 0.35$)
&   8.3 \; \scriptsize($ 0.04$) &  59.9 \scriptsize($ 0.05$) &   71.2 \scriptsize($ 2.11$)
&   3.7 \; \scriptsize($ 0.01$) &  54.5 \scriptsize($ 0.07$) &   40.4 \scriptsize($ 0.14$)
\\
20 & 2
&  28.5 \scriptsize($ 0.23$) & 153.0 \scriptsize($ 1.57$) & 1152.4 \scriptsize($ 9.80$)
&  11.6 \scriptsize($ 0.08$) &  72.1 \scriptsize($ 0.16$) &  102.4 \scriptsize($ 0.72$)
&   5.3 \; \scriptsize($ 0.03$) &  56.4 \scriptsize($ 0.16$) &   44.9 \scriptsize($ 0.10$)
\\
20 & 3
&  20.8 \scriptsize($ 0.06$) & 102.9 \scriptsize($ 1.08$) &  956.4 \scriptsize($ 8.11$)
&   8.3 \; \scriptsize($ 0.06$) &  64.7 \scriptsize($ 0.08$) &   92.6 \scriptsize($ 0.48$)
&   3.9 \; \scriptsize($ 0.02$) &  55.1 \scriptsize($ 0.05$) &   41.7 \scriptsize($ 0.28$)
\\
\bottomrule
\end{tabular}%
}
\end{table}

\subsection{Sensitivity analysis to $\rho_{\min}$}\label{app:rho_sensitivity}

\begin{figure}[!h]
    \centering
    \includegraphics[width=0.7\linewidth]{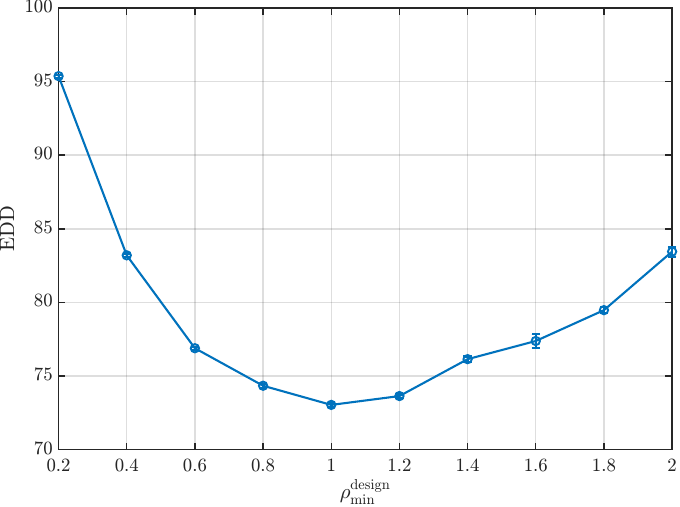}
    \caption{Sensitivity of MRS-C detection delay to the design lower bound $\rho_{\min}^{\mathrm{design}}$ under $k=10$, $d=3$, $\sigma^2=1$, $\Lambda=I_d$, and $w=50$. Points show mean detection delay over five batches of $10^4$ post-change runs, with error bars indicating one standard deviation.}
    \label{fig:rho_min_sensitivity}
\end{figure}

We also examine the sensitivity of MRS-C to the design lower bound
$\rho_{\min}$. We fix $k=10$, $d=3$, $\sigma^2=1$, $\Lambda = I_d$ and $w=50$, and vary $\rho_{\min}^{\mathrm{design}}\in\{0.2,0.4,\dots,2.0\}$. 
The threshold is recalibrated for each design value to maintain the same target ARL of 5000.

The resulting EDD is non-monotone in the design lower bound. As in Figure \ref{fig:rho_min_sensitivity},
the EDD decreases until the true $\rho_{\min} = 1.0$, and then increases.
Thus, overly conservative and overly aggressive choices can both increase delay,
while a value near the true value performs best, as expected.

\subsection{Hybrid detector: MRS-C and LESC}\label{app:hybrid}

\begin{figure}[!h]
\centering
\includegraphics[width=\textwidth]{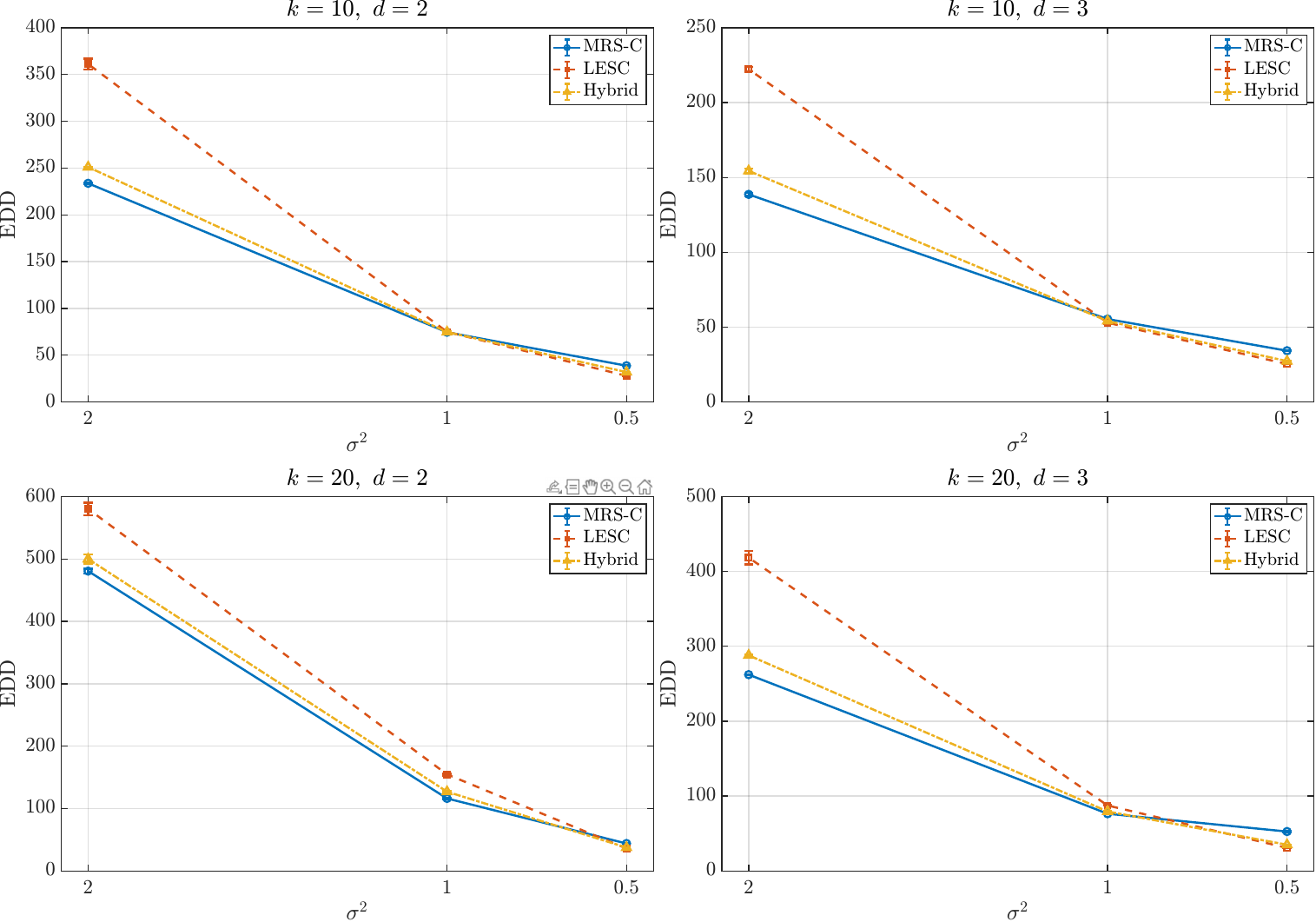}
\caption{EDD comparison of MRS-C, LESC, and the hybrid detector under the
uniform-spike model for $k\in\{10,20\}$, $d\in\{2,3\}$, and
$\sigma^2\in\{2,1,0.5\}$. All procedures are calibrated to the same target ARL; the
hybrid rule is jointly calibrated because either component chart can trigger an alarm.}
\label{fig:hybrid}
\end{figure}

We also evaluate the hybrid rule discussed in Remark~\ref{rem:hybrid}. The hybrid detector runs MRS-C and LESC in parallel and stops at the first alarm,
\[
T_{\mathrm{hybrid}}=\min\{T^{\mathrm{sub}},T_{\mathrm{LESC}}\}.
\]
For a fair comparison, the thresholds of the hybrid rule are jointly calibrated so that the combined stopping rule attains the same target ARL as the standalone procedures.

Figure \ref{fig:hybrid} compares the EDDs of MRS-C, LESC, and the hybrid detector under the uniform-spike model for $k\in\{10,20\}$, $d\in\{2,3\}$, and $\sigma^2\in\{2,1,0.5\}$.
The threshold for each method is calibrated to attain an ARL of 5000.
The results show that the hybrid detector does not uniformly dominate the better standalone method.
In the low-SNR setting $(\sigma^2=2)$, MRS-C is typically the fastest, while the hybrid rule is slightly slower because of the joint ARL calibration penalty.
In the high-SNR setting $(\sigma^2=0.5)$, LESC is typically faster, and the hybrid rule remains close to LESC but does not consistently improve upon it.
Thus, the hybrid detector should be viewed as a practical hedge when the signal strength is unknown, rather than as a uniformly superior procedure.

\section{Network-based reference methods}
\label{app:network_baselines}

This section gives the definitions needed for the network-based comparisons in Section~\ref{subsubsec:milling}. Further algorithmic details are deferred to the cited papers. For this application, the normalized nodal features define the symmetric Gram-affinity matrix
\[
A_t=\widetilde X_t^T\widetilde X_t\in\Real^{n\times n}.
\]
Agent identities are retained across time, providing the one-to-one vertex correspondence required by the longitudinal network methods below.

\subsection{Spectral CUSUM \citep{zhang2023spectral}}
\label{app:spectral-cusum}

Spectral CUSUM of \citet{zhang2023spectral} begins with a vector $\zeta_t\in\Real^n$ containing one scalar signal per node. For a future-window length $w_{\mathrm{SC}}$, it uses
\[
\widehat G_t^{\mathrm{SC}}=\frac{1}{w_{\mathrm{SC}}}
\sum_{s=1}^{w_{\mathrm{SC}}}\zeta_{t+s}\zeta_{t+s}^T.
\]
If $(\widehat\lambda_{tj}^{\mathrm{SC}},\widehat e_{tj}^{\mathrm{SC}})$, $j=1,\ldots,m_{\mathrm{SC}}$, are the $m_{\mathrm{SC}}$ smallest eigenpairs of $\widehat G_t^{\mathrm{SC}}$, its scalar spectral score is
\[
g_t^{\mathrm{SC}}=\sum_{j=1}^{m_{\mathrm{SC}}}
\frac{\{(\widehat e_{tj}^{\mathrm{SC}})^T\zeta_t\}^2}
{\widehat\lambda_{tj}^{\mathrm{SC}}}.
\]
With drift $\Delta_{\mathrm{SC}}$ and threshold $b_{\mathrm{SC}}$, the recursion in \citet[Algorithm~1]{zhang2023spectral} is
\[
S_t^{\mathrm{SC}}=(S_{t-1}^{\mathrm{SC}})_+-g_t^{\mathrm{SC}}+\Delta_{\mathrm{SC}},
\quad S_0^{\mathrm{SC}}=0,
\]
and the corresponding alarm time is
\[
T_{\mathrm{SC}}=\inf\{t>0:S_t^{\mathrm{SC}}\geq b_{\mathrm{SC}}\} + w_{\mathrm{SC}}.
\]

\subsection{Iso-mirror \citep{chen2023discovering}}
\label{app:iso-mirror}

Iso-mirror is retrospective: in Figure~\ref{fig:results}, it uses all affinity matrices on the selected physical-time grid $\mathcal I=\{780 \mathrm{s},780.1 \mathrm{s},\ldots,820 \mathrm{s}\}$. This is an adaptation of the aligned-adjacency-matrix construction of \citet{chen2023discovering} to the Gram-affinity matrices above. Define the rank-$r_{\mathrm{iso}}$ adjacency spectral embedding by
\[
\operatorname{ASE}_{r_{\mathrm{iso}}}(A)
=\bigl[\sqrt{\lambda_1(A)}u_1(A),\ldots,
\sqrt{\lambda_{r_{\mathrm{iso}}}(A)}u_{r_{\mathrm{iso}}}(A)\bigr],
\]
where $\lambda_j(A)$ and $u_j(A)$ are the leading eigenpairs of $A$. For two times $s,t\in\mathcal I$, the estimated network dissimilarity is
\[
\widehat d_{st}=\frac{1}{\sqrt n}
\min_{W^TW=I_{r_{\mathrm{iso}}}}
\left\|\operatorname{ASE}_{r_{\mathrm{iso}}}(A_s)
-\operatorname{ASE}_{r_{\mathrm{iso}}}(A_t)W\right\|_2,
\]
where $\|\cdot\|_2$ is the matrix spectral norm. The orthogonal alignment removes the non-identifiability of the individual spectral embeddings.

Let $\widehat D=(\widehat d_{st})$. Classical multidimensional scaling applied to $\widehat D$ produces the estimated Euclidean mirror; one-dimensional ISOMAP applied to those mirror points then produces the scalar iso-mirror coordinate $\widehat\psi_t$ (up to an immaterial translation and reflection). For our retrospective summary, we use the continuous one-break segmented regression described by \citet{chen2023discovering},
\[
\widehat\tau_{\mathrm{iso}}
=\arg\min_{\tau\in(780,820)}\ \min_{\alpha,\beta,\gamma}
\sum_{t\in\mathcal I}\left\{\widehat\psi_t-\alpha-\beta t-\gamma(t-\tau)_+\right\}^2.
\]
The fitted breakpoint $\widehat\tau_{\mathrm{iso}}$ is the retrospective transition estimate used in the comparison. Further details on the eigenspace computation, mirror construction, neighborhood graph in ISOMAP, and segmented-regression fitting are deferred to \citet{chen2023discovering}.

\bibliographystyle{tfcad}
\bibliography{references}

\end{document}